# High-pressure lattice dynamics in bulk single-crystal BaWO$_4$

F.J. Manjón,[1,2,*] D. Errandonea,[1] N. Garro,[1] J. Pellicer-Porres,[1] P. Rodríguez-Hernández,[3] S. Radescu,[3] J. López-Solano,[3] A. Mujica,[3] and A. Muñoz[3]

[1] Dpto. de Física Aplicada i Institut de Ciència de Materials de la Universitat de València, 46100 Burjassot (València), Spain
[2] Dpto. de Física Aplicada, Universitat Politècnica de València, 46022 València, Spain
[3] Dpto. de Física Fundamental II, Universidad de La Laguna, 38205 La Laguna, Tenerife, Spain

**Abstract.** Room-temperature Raman scattering has been measured in barium tungstate (BaWO$_4$) up to 16 GPa. We report the pressure dependence of all the Raman-active first-order phonons of the tetragonal scheelite phase (BaWO$_4$-I, space group I4$_1$/$a$), which is stable at normal conditions. As pressure increases the Raman spectrum undergoes significant changes around 6.9 GPa due to the onset of the structural phase transition to the monoclinic BaWO$_4$-II phase (space group P2$_1$/$n$). This transition is only completed above 9.5 GPa. A further change in the spectrum is observed at 7.5 GPa related to a scheelite-to-fergusonite transition. The scheelite, BaWO$_4$-II, and fergusonite phases coexist up to 9.0 GPa due to the sluggishness of the I→II phase transition. Further to the experimental study, we have performed *ab initio* lattice dynamics calculations that have greatly helped us in assigning and discussing the pressure behaviour of the observed Raman modes of the three phases.

---

[*] Corresponding author with present address at UPV. E-mail address: fjmanjon@fis.upv.es
Tel.: + 34 96 387 52 87, Fax: + 34 96 387 71 89

**I. Introduction**

Alkaline-earth tungstates are key compounds in the improvement of instrumentation for low-background particle physics experiments. The development of cryogenic phonon-scintillation detectors **[1]** could provide simultaneous measurement of phonon and scintillation signals, making possible event-type discrimination measurements for the search of rare events **[2-4]**. For this application, the understanding of the lattice dynamics of these compounds is important in order to improve the knowledge and control of their thermal and optical properties.

Barium tungstate ($BaWO_4$) is the heaviest member of the family of the alkaline-earth tungstates. Like many other $ABX_4$–type compounds, $BaWO_4$ crystallizes at ambient conditions in the tetragonal scheelite-type structure (space group [SG]: $I4_1/a$, No. 88, Z = 4) **[5]**. $BaWO_4$ has also been observed at ambient conditions in a metastable monoclinic polymorph named $BaWO_4$-II (SG: $P2_1/n$, No. 14, Z = 8), which can be obtained from the scheelite phase after a high-pressure high-temperature treatment **[6]**. This monoclinic structure was thus considered as a potential stable structure for tungstates at moderately high pressures and room temperature (RT) **[7,8]**.

High-pressure Raman studies of alkaline-earth tungstates ($CaWO_4$, $SrWO_4$, $BaWO_4$) **[7]** and $PbWO_4$ **[8]** were performed up to 9 GPa by Jayaraman *et al*. motivated by previous studies on $CaWO_4$ **[9]**, and on $SrWO_4$ **[10]** up to 4 GPa. These authors found that $BaWO_4$ underwent a phase transition near 6.5 GPa and suggested that the high-pressure phase could have octahedral W coordination as in the $HgWO_4$–type structure (SG: $C2/c$, No. 15, Z = 4) **[7]**. More recent Raman studies on $CaWO_4$ and $SrWO_4$ have also found a phase transition above 10 GPa **[11,12]**, but concluded that the fourfold W coordination in the scheelite phase was retained in the high-pressure phase, which is consistent with a transition towards the monoclinic M-fergusonite structure (SG: $I2/a$, No. 15, Z = 4), hereafter called simply fergusonite. Indeed, the RT pressure-driven scheelite-to-ferguson ite transition has been recently observed in angle dispersive powder x-ray diffraction (ADXRD) and near-edge x-ray absorption structure (XANES) experiments under pressure on several alkaline-earth tungstates **[13-17]**. On the basis of these structural studies, it can be concluded that the high-pressure Raman spectra of $CaWO_4$ **[11]** and $SrWO_4$ **[12]** above 12 GPa and up to 20 GPa can be undoubtedly assigned to the fergusonite phase. In $BaWO_4$, the phase transition to the fergusonite phase has been reported to occur near 7 GPa **[14,17]**. However, the Raman spectra of the high-pressure phase of $BaWO_4$ previously measured above 6.5 GPa **[7]** does not seem to resemble those of the high-pressure phase of $CaWO_4$ and $SrWO_4$ **[11,12]**. In addition, the comparison of the Raman spectra of $BaWO_4$ above 6.5 GPa **[7]** and the results reported for $PbWO_4$ above 4.5 GPa **[8]** show significant differences. These differences led to Jayaraman *et al*. to conclude that the high-pressure phase of $BaWO_4$ likely was of the $HgWO_4$-type and different to that of the high-pressure phase of $PbWO_4$ **[8]**.

In order to shed light on the high-pressure phase transitions in these important and complex compounds, Errandonea *et al*. have recently conducted ADXRD and XANES measurements under pressure on $CaWO_4$, $SrWO_4$, $BaWO_4$, and $PbWO_4$ **[16,17]**. They observed the scheelite-to-fergusonite transition in the four tungstates investigated and suggested the existence of other pressure-induced transitions. In particular, a scheelite-to-$BaWO_4$-II phase transition was suggested for $BaWO_4$ **[17]**. Their *ab initio* calculations established that the scheelite phase in $BaWO_4$ was unstable respect to the $BaWO_4$-II and fergusonite phases at 5 and 7.5 GPa, respectively **[17]**. However, the structural ADXRD measurements of Errandonea *et al*. performed in powder samples of $BaWO_4$ evidenced the fergusonite phase before the $BaWO_4$-II phase. They interpreted their findings in terms of a large activation barrier associated to the scheelite-to-$BaWO_4$-II phase transition **[17]**. In fact, the similarity between the scheelite and ferguson ite phases made difficult even the observation of the ferguson ite phase, which could be only inferred from very subtle diffraction peaks in the ADXRD spectra **[17]**. The pressure for the first phase transition in BaWO$_4$ obtained from ADXRD measurements in **Ref. 17** (7.1 GPa) is in reasonable agreement with the previous ADXRD measurements of Panchal *et al*. (7.0 GPa) **[14]**, and with previous Raman measurements of Jayaraman *et al*. (6.5 GPa) **[7]**. Furthermore, it also agrees with an estimate of the transition pressure based on a systematic study of scheelite-related compounds (5.9 GPa) **[18].**

The previous Raman work on BaWO$_4$ of Jayaraman *et al*. **[7]** neither report the second phase transition in BaWO$_4$, nor discussed the nature of the modes of the first high-pressure phase observed **[7]**. In this study, we present a RT Raman study of single crystals of BaWO$_4$ up to 16 GPa as part of our project to study the stability of scheelite-structured orthotungstates. The Raman measurements are complemented with *ab initio* lattice dynamics calculations in order to help us in the assignment and discussion of the behavior of the zone-center phonons in the different structural phases. Our objective is to shed some light into the complex high-pressure phase diagram of these interesting materials and give a comprehensive description of the behavior of the Raman modes of BaWO$_4$ (in both its low and high-pressure phases) and their related compounds. This can be done with Raman scattering which is a subtle probe capable of distinguishing small traces of various local phases coexisting in a compound and thus has the ability to detect pressure-induced phase transitions at lower pressures than ADXRD and XANES measurements **[19]**. We will show that the Raman spectra of BaWO$_4$ under high pressure phases reported by Jayaraman *et al*. **[7]** can be completely understood on the light of the present work.

**II. Experimental details**

The BaWO$_4$ samples used in this study were obtained from scheelite-type bulk single crystals grown with the Czochralski method. The starting raw powders had 5N purity [20]. Small platelets (100μm x 100μm x 30μm) were cleaved from these crystals along the {101} natural cleavage plane [21] and inserted in a diamond-anvil cell (DAC). The pressure-transmitting medium used was a 4:1 methanol-ethanol mixture and the pressure was determined by calibration with the ruby photoluminescence [22]. RT Raman experiments were performed in backscattering geometry using the 488 nm line of an Ar$^+$-ion laser with a power of less than 100 mW before the DAC. Dispersed light was analysed with a Jobin-Yvon T64000 triple spectrometer equipped with a confocal microscope in combination with a liquid nitrogen-cooled multi-channel CCD detector. Spectral resolution was better than 1 cm$^{-1}$ and Ar and He plasma lines were used to calibrate the Raman and photoluminescence spectra.

**III. *Ab initio* lattice dynamics calculations.**

Along with the experimental data of our Raman study we will also present results of a theoretical *ab initio* calculation of the phonon modes of the scheelite, fergusonite, and BaWO$_4$-II phases at the zone center (Γ point). Besides its inherent interest for the sake of comparison with the experiment, the results of this theoretical study have allowed us to assign the observed modes among the three structures in their pressure region of coexistence as well as providing information about the symmetry of the modes and polarization vectors which is not readily accessible in the present experiment. All the calculations were done within the framework of the density functional theory (DFT) using the Vienna *ab initio* simulation package (VASP) of which a detailed account can be found in **Ref. 23** and references therein. The exchange and correlation energy was initially taken in the local density approximation (LDA) for tests and afterwards taken in the generalized gradient approximation (GGA) according to Perdew-Burke-Ernzerhof (GGA-PBE) for a more exact calculation. The projector-augmented wave (PAW) scheme **[24]** was adopted and the semicore 5s and 5p electrons of Ba were dealt with explicitly in the calculations. The set of plane waves used extended up to a kinetic energy cutoff of 910 eV. This large cutoff was required to deal with the O atoms within the PAW scheme. The Monkhorst-Pack grid used for Brillouin-zone integrations ensured highly converged results (to about 1 meV per formula unit). At each selected volume, the structures were fully relaxed to their equilibrium configuration through the calculation of the forces on atoms and the stress tensor – see **Ref. 17**. In the relaxed equilibrium configuration, the forces are less than 0.002 eV/A and the deviation of the stress tensor from a diagonal hydrostatic form is less than 1 kbar. The highly converged results on forces are required for the calculation of the dynamical matrix using the direct force constant approach (or supercell method) **[25]**. The construction of the dynamical matrix at the Γ point is particularly simple and involves separate calculations of the forces in which a fixed displacement from the equilibrium configuration of the atoms within the *primitive* unit cell is considered. Symmetry aids by reducing the number of such independent distortions – to a mere 36 independent displacements in the $BaWO_4$-II phase, which is the more demanding of all the structures considered. Diagonalization of the dynamical matrix provides both the frequencies of the normal modes and their polarization vectors.

**IV. Results and discussion**

**A. Low-pressure phase: Scheelite structure**

The scheelite structure of $BaWO_4$, adopted by $BaWO_4$ at low pressures, is centrosymmetric and has space group $I4_1/a$ ($C^6_{4h}$ in Schoenflies notation) with four formula units per body centered unit cell. The Ba and W atoms occupy $S_4$ sites and the sixteen oxygen atoms are on $C_1$ sites. Group theoretical considerations lead to the following vibrational representation at the $\Gamma$ point for scheelite-$BaWO_4$ [26], in standard notation:

$$\Gamma = (3A_g + 3B_u) + (5B_g + 5A_u) + (5E_g + 5E_u) \qquad (1)$$

A and B modes are non-degenerate, whereas the E modes are doubly degenerate. The subindex g and u stand for *even* and *odd*, respectively, and indicate parity under inversion in centrosymmetric crystals. One $A_u$ and one $E_u$ correspond to zero frequency acoustic modes, the rest are optic modes. The pairs of species enclosed in parenthesis arise from the same motion of the $BaWO_4$ molecule. In scheelites, the first member of the pairs (g) is Raman-active and the second member (u) is infrared (IR)-active, except for the $B_u$ silent modes that are not IR-active. Consequently, we expect 13 zone-center Raman-active modes in scheelite-$BaWO_4$:

$$\Gamma = 3A_g + 5B_g + 5E_g \qquad (2)$$

The vibrational spectra of $ABO_4$ scheelites have been interpreted in terms of modes of the $BO_4$ tetrahedra [27], inasmuch as these tetrahedra remain relatively undistorted and independent units in the scheelite phase. Thus, the scheelite modes have been classified either as *internal*, when the $BO_4$ center of mass does not move, or as *external*, when they imply movement of $BO_4$ tetrahedra as rigid units. The framework for this description can be readily established as follows. The $BO_4$ molecule has 5 atoms and thus supports 15 modes of motion, including 3 pure rotations and 3 pure translations. The $T_d$ symmetry of the $BO_4$ molecule leads to 6 zone-center modes following the decomposition:

$$\Gamma_{15} = A_1 + E + F_1 + 3F_2 \qquad (3)$$

where $F_1$ corresponds to a pure rotation (noted as R in the literature), and one of the $F_2$ is a pure translation (noted as T). R and T modes are related to the external modes observed in ABO$_4$ scheelites. The scheelite modes derived from the tetrahedral A$_1$, E, and 2F$_2$ modes are usually named as $\nu_1$, $\nu_2$, $\nu_3$ and $\nu_4$ **[28]**, respectively, and are considered as internal modes of the BO$_4$ tetrahedra in ABO$_4$ scheelites.

In scheelite-BaWO$_4$, the W site has S$_4$ symmetry instead of the full T$_d$ tetrahedral symmetry of the WO$_4$ group. The reduction of the T$_d$ representation under the S$_4$ symmetry of the WO$_4$ site in the BaWO$_4$ scheelite lattice, and the correlation of the S$_4$ site symmetry to the C$_{4h}$ point group corresponding to $\Gamma$, leads to the following 13 zone-center Raman-active modes:

$$\Gamma = \nu_1(A_g) + \nu_2(A_g) + \nu_2(B_g) + \nu_3(B_g) + \nu_3(E_g) + \nu_4(B_g) + \nu_4(E_g)$$
$$+ R(A_g) + R(E_g) + 2T(B_g) + 2T(E_g) \qquad (4)$$

The translational modes are the lowest in frequency, the internal modes are the highest in frequency, and the frequencies of the rotational modes are between those of the translational and the internal modes. To the best of our knowledge only twelve of the thirteen modes in scheelite-BaWO$_4$ are known **[7]**, with the internal mode $\nu_2(A_g)$ being the only unknown one. However, only the pressure dependence of nine of the twelve modes has been previously reported **[7]**. In order to assign the different Raman modes of scheelite-BaWO$_4$ we have followed the notation of Liegeois-Duyckaerts and Tarte **[29]**.

**Figure 1** shows the RT Raman spectra of scheelite-BaWO$_4$ at different pressures up to 7.5 GPa. The Raman spectra should correspond to a mixture of polarizations perpendicular and parallel to the c-axis because of our sample orientation. In **Fig. 1**, we have marked at the bottom the *ab initio* calculated frequencies of the Raman modes in scheelite-BaWO$_4$ at 1 atm. The Raman spectrum of the scheelite phase is dominated by the $\nu_1(A_g)$ and $\nu_2(B_g)$ modes near 926 cm$^{-1}$ and 332 cm$^{-1}$ at 1 atm, respectively. At first sight, one can distinguish twelve out of the thirteen modes in the Raman spectrum of scheelite-BaWO$_4$ up to 6.9 GPa. The $\nu_2(A_g)$ mode is the only mode that can not be clearly observed. However, the mode located near 332 cm$^{-1}$ at 1 atm, which was initially attributed to the $\nu_2(B_g)$ mode, is indeed a double mode. This can be inferred from the asymmetry of the peak after a careful inspection of the experimental spectra made on the basis of the closeness of our theoretically calculated $\nu_2$ modes (see **Fig. 1** and **Table I**). These experimental observations and theoretical calculations are in agreement with suggestions of previous authors that considered that both the two $\nu_2$ and the two $\nu_4$ modes were very close in frequency with the splitting between the two $\nu_2$ modes being smaller than that of the two $\nu_4$ modes, and with the intensities of the $\nu_2$ modes being in general considerably larger than the $\nu_4$ ones **[30,31]**.

A detail of the Raman spectra near the double mode at several pressures is shown in **Fig. 2**. The spectra at different pressures have been shifted in frequency in order to bring the $\nu_2(B_g)$ mode into coincidence, so as to show the relative shift of the low-frequency shoulder with respect to the main mode as a function of pressure. A marked broadening of the $\nu_2(B_g)$ mode in its low-frequency tail is observed in **Fig. 2** with increasing pressure. We think that this broadening is the marked feature of the presence of the $\nu_2(A_g)$ mode at the low frequency side of the $\nu_2(B_g)$ mode. The assignment of the $\nu_2(A_g)$ mode is based on the ordering of the two modes and on the slightly smaller pressure coefficient of the $\nu_2(A_g)$ mode than that of the $\nu_2(B_g)$ mode, as indicated by our *ab initio* lattice dynamics calculations. The smaller pressure coefficient of the $\nu_2(A_g)$ mode leads to an apparent broadening of the low-frequency side of the $\nu_2(B_g)$ mode when both modes separate each other. The reason for the different pressure coefficients of the two $\nu_2$ modes is that the $\nu_2(A_g)$ mode only implies the movement of O atoms while the $\nu_2(B_g)$ mode introduces an additional displacement of Ba along the c-axis and thus it should be more sensitive to pressure **[27]**. In order to improve the estimation of the pressure coefficient of the $\nu_2(A_g)$ mode with respect to the $\nu_2(B_g)$ mode, we have performed the numerical second derivative of the Raman spectra (see **inset of Fig. 2**). This technique allows us to track small changes in the Raman profiles **[32]**. In this way, we have located the low-frequency $\nu_2(A_g)$ mode at 331 cm$^{-1}$ at 1 atm.

**Figure 3** shows the pressure (P) dependence of all the frequency of all the Raman modes (solid circles) of scheelite-BaWO$_4$ up to 7 GPa. **Table I** summarizes the frequencies ($\omega$) of all the scheelite Raman modes, their pressure coefficients (d$\omega$/dP), and Grüneisen parameters ($\gamma = B_0/\omega \cdot d\omega/dP$, with $B_0 = 52$ GPa being the scheelite bulk modulus at ambient pressure **[17]**). In **Table I**, we also compare the experimental results for scheelite-BaWO$_4$ to those from our *ab initio* lattice dynamics calculations, and to those from previous experimental results for other alkaline-earth tungstates **[7,11,12]**. For completeness, **Table II** summarizes our calculated frequencies and frequency pressure coefficients for the IR modes of scheelite-BaWO$_4$ at 1 atm. The frequencies of the IR modes compare well with those reported experimentally **[33-35]**. Note that the polar IR modes $A_u$ and $E_u$ split into a TO and a LO mode. Both modes are given in **Table II** after **Ref. 33**, but only the TO mode can be compared to our calculated values.

Our measured frequencies, pressure coefficients and Grüneisen parameters in scheelite-BaWO$_4$ agree reasonably with our calculated values of these magnitudes at the same pressure. Our results also agree with the values reported by Jayaraman *et al.* **[7]** once their Grüneisen parameters are corrected using the more recently measured bulk modulus values (52 **[17]** and 57 GPa **[14]**). The only deviation with respect to Jayaraman *et al.* corresponds to the pressure coefficient of the T($B_g$) mode. While these authors reported a zero pressure coefficient for this mode located at 104 cm$^{-1}$ at 1 atm **[7]**, we have measured a relatively high positive pressure coefficient (3.3 cm$^{-1}$/GPa) for this mode, which we have located at 101 cm$^{-1}$ at 1 atm in good agreement with the previously reported value (102 cm$^{-1}$) **[29]**. We think that a possible reason for the discrepancy in the pressure coefficient of this mode between Jayaraman *et al.* and us is that the mode measured at 104 cm$^{-1}$ by Jayaraman *et al.* could be an Ar-ion plasma line instead of the T($B_g$) mode. In fact, we have used a plasma line at 104 cm$^{-1}$ with respect to the 488 nm laser line to calibrate our Raman spectra [see vertical dashed line in **Figs. 1, 4 and 5(a)**].

**B. High-pressure phases**

**Figure 4** shows the Raman spectra of BaWO$_4$ at different pressures ranging from 6.9 up to 16 GPa. At 6.9 GPa many Raman modes of the scheelite phase still can be observed together with some new Raman peaks that do not correspond to the scheelite phase (see double arrows in **Fig. 4**). Therefore, we take this value as the pressure of the onset of the first phase transition (see dotted line in **Fig. 3**). In this sense, we must note that on decreasing pressure from 16 GPa the scheelite phase of BaWO$_4$ was recovered below 3 GPa after a considerable hysteresis. It can be seen that, apart from the new peaks appearing at 6.9 GPa, there are new Raman peaks appearing at 7.5 GPa. We will show in the following that this pressure marks the onset of a second phase transition (see also dotted line in **Fig. 3**). In order to distinguish between modes of these two phase transitions, we have marked the positions of the Raman peaks appearing at 6.9 GPa (both at 6.9 and 7.5 GPa) with double arrows in **Fig. 4**. The Raman modes of the two new phases that appear at 6.9 and 7.5 GPa coexist with those of the scheelite phase up to 9.0 GPa since evidence of the presence of the $\nu_1$($A_g$) mode of the scheelite structure is observed up to this pressure.

The Raman spectra of BaWO$_4$ above 7.5 GPa are completely different to that of the scheelite phase. They are dominated by three high-frequency modes and a broad band near 350 cm$^{-1}$ showing a number of peaks. The most striking features of the spectrum at 7.5 GPa are: 1) the large number of modes observed, as compared to the scheelite phase, which is even larger than the number expected for the fergusonite phase, as will be discussed below. Some of the new modes are located between 400 and 800 cm-1, a previously deserted range; 2) the sudden broadening of the scheelite $\nu_1(A_g)$ mode; 3) the appearance of two modes near the scheelite $\nu_1(A_g)$ mode; 4) the splitting of the scheelite $\nu_3$ modes near 800 cm$^{-1}$ in a number of weak modes; and 5) the appearance of several modes near 350 cm$^{-1}$. In particular, a mode located at 338 cm$^{-1}$ appears at 7.5 GPa and disappears already at 8.2 GPa.

In the following discussion we will show that the new Raman peaks appearing at 6.9 GPa correspond to the BaWO$_4$-II phase, while the new Raman modes appearing at 7.5 GPa correspond to the fergusonite phase. The assignment of the Raman modes appearing above 6.9 GPa to these two different phases is based on the classification of these modes into two types of modes: 1) modes that appear at 7.5 GPa and disappear above 9.0 GPa; 2) modes that can be followed in pressure up to almost 16 GPa, some of them already appearing at 6.9 GPa. The different pressure behavior of these two types of modes leads us to believe that between 7.5 and 9.0 GPa we have a mixture of two high-pressure phases with the scheelite one. The different pressure behavior of the two types of new modes is most clearly observed in the three strong high-frequency stretching modes appearing in the Raman spectra between 900 and 950 cm$^{-1}$ above 7.5 GPa (see **Fig. 4**). It can be seen that the broad mode near 940 cm$^{-1}$ present at 7.5 and 8.2 GPa disappears at 9 GPa (see arrow in **Fig. 4**) while the other two strong high-frequency stretching modes remain up to the highest pressure attained in the experiment. Therefore, the strong mode near 940 cm$^{-1}$ is assigned to the fergusonite phase while the other two strong modes are assigned to the BaWO$_4$-II phase. A similar decrease in intensity between 7.5 and 9 GPa can be clearly observed in three modes near 830 cm$^{-1}$, in the lowest frequency mode near 40 cm$^{-1}$, and in the sudden appearance and disappearance of the mode located at 338 cm$^{-1}$ at 7.5 GPa (see arrows in **Fig. 4**). All these modes appearing at 7.5 GPa and disappearing at most at 9 GPa are assigned to the fergusonite phase. In sections B.1 and B.2 we will discuss the nature of the different modes assigned to the high-pressure phases.

**B.1. Fergusonite phase**

In order to study the nature of the Raman modes of the high-pressure phases let us begin with the modes expected for the fergusonite phase since the scheelite-to-fergusonite transition has been observed by previous powder ADXRD and XANES measurements **[13-17]**. Besides, we can compare our high-pressure Raman spectra for $BaWO_4$ with those available in the literature for $CaWO_4$, $SrWO_4$ and $BaWO_4$ **[11,12,7]**. In this respect, it is mandatory to say that the Raman modes of $CaWO_4$ and $SrWO_4$ reported between 12 and 20 GPa **[11,12]** correspond to the fergusonite phase since this phase is the only one found in ADXRD and XANES measurements in that pressure range **[16]**.

The body-centered monoclinic fergusonite structure is centrosymmetric and has space group I2/$a$ ($C_{2h}^6$) with four formula units per conventional unit cell (i.e., two formula units per primitive unit cell). Group theoretical considerations indicate that fergusonite-$BaWO_4$ should have 36 vibrational modes at the zone center with the following mechanical representation **[26]**:

$$\Gamma = 8A_g + 8A_u + 10B_g + 10B_u \qquad (5)$$

with the 18 gerade (g) modes being Raman active and the 18 ungerade (u) modes being IR active. The 18 Raman-active modes derive from the reduction of the tetragonal $C_{4h}$ symmetry of the scheelite structure to the monoclinic $C_{2h}$ symmetry of the fergusonite structure. In particular, every $A_g$ and every $B_g$ scheelite mode transforms into an $A_g$ mode of the monoclinic symmetry, while every doubly degenerate $E_g$ scheelite mode transforms into two $B_g$ modes of the monoclinic symmetry. Therefore, we expect 18 zone-center Raman-active modes after the scheelite-to-fergusonite phase transition.

Above 6.9 GPa many Raman modes of the scheelite structure have either weakened considerably or disappeared, and the number of new modes measured exceeded the number of modes expected for the fergusonite structure. These results suggest that either the high-pressure phase is not fergusonite or that there is a mixture of phases with one of them being fergusonite, as concluded from ADXRD measurements **[17]**. The above classification of the two classes of Raman modes appearing above 6.9 GPa and 7.5 GPa made clear that there is a mixture of phases between 7.5 and 9.0 GPa. Therefore, taking into account first of all that recent ADXRD measurements observed a fergusonite phase above 7.1 GPa **[14,17]**, that the results of *ab initio* total-energy calculations showed the larger stability of the $BaWO_4$-II phase respect to the fergusonite phase at high pressures **[17]**, and finally that the number of modes appearing at 7.5 GPa and disappearing near 9 GPa is around twelve, we attribute these modes to the fergusonite phase, whereas the modes that appear at 6.9 GPa and last up to almost 16 GPa are attributed to the $BaWO_4$-II phase.

**Figures 5(a) and (b)** show details of the Raman spectra of $BaWO_4$ between 6.9 and 9 GPa in two different wavelength ranges. The *ab initio* calculated frequencies of the eighteen fergusonite Raman modes at 8.2 GPa are marked at the bottom of **Figs. 5(a) and (b)**. In these figures one can observe the appearance of the modes assigned to the fergusonite phase (marked by asterisks) at 7.5 GPa and their fading around 9 GPa. **Fig. 3** shows the frequency pressure dependence of the fergusonite modes (blank squares), and **Table III** summarizes the experimental Raman frequencies, pressure coefficients and relative frequency pressure coefficients of the observed modes of fergusonite-$BaWO_4$ at 7.5 GPa.

The structural similarities between the scheelite and fergusonite phases give clues for the identification and assignment of the modes in the fergusonite phase. Since the fergusonite phase retains the tetrahedral coordination of W (see section C), with only a slight distortion of the WO$_4$ tetrahedra, we expect small changes in the frequencies of the internal modes. However, the change in lattice parameters between the two phases; in particular, the shear of alternate (100) cation planes and the small distortion of the beta angle, lead to expect larger changes in the frequencies of the external modes. On the other hand, it can be observed that the most intense Raman modes in the ferrgusonite structure derive from the most intense modes in the scheelite phase. This means that the strong $\nu_1$ and $\nu_2$ modes in scheelite-BaWO$_4$ are transformed into strong modes in the ferrgusonite phase. This observation is coherent, for instance, with the fact that the $A_g$ Raman modes in the scheelite phases are only due to O vibrations in the 16f Wyckoff positions, and also O vibrations give the main contribution to $A_g$ modes in the ferrgusonite phase **[36]**. Therefore, the Raman modes of the scheelite and ferrgusonite phases are not only related but they seem to exhibit similar scattering cross sections. We have also taken this feature as a key for the assignment of the nature of the different Raman modes in the ferrgusonite phase.

Additional support for the location and assignment of the ferrgusonite modes is provided by our *ab initio* lattice dynamics calculations [see marks at the bottom of **Figs. 5(a) and (b)**]. **Table III** summarizes the calculated Raman mode frequencies, their pressure coefficients, and their symmetry for ferrgusonite-BaWO$_4$ at 8.2 GPa. For completeness, we also report in **Table IV** the frequencies and the frequency pressure coefficients of the calculated IR-active modes of ferrgusonite-BaWO$_4$ at 8.2 GPa. A major conclusion drawn from our calculations is that there is a phonon gap between 400 and 800 cm$^{-1}$ in ferrgusonite-BaWO$_4$ similar to the one observed in scheelite-BaWO$_4$. This result not only allows us to confirm that the ferrgusonite phase retains the tetrahedral W-O coordination of the scheelite phase (see section C), but it also leads us to conclude that the Raman modes observed at high pressures in this phonon gap belong to a phase other than the scheelite or the ferrgusonite. As regards to the nature of the ferrugsonite Raman modes, it is difficult to assign the symmetry of all the modes in this structure despite of having the aid of the *ab initio* calculations because we have only found 12 of the 18 Raman-active modes of the ferrugsonite phase, and also because the small pressure range in which the ferrugsonite Raman modes are observed (7.5-9.0 GPa) limited the accuracy of the measured pressure coefficients.

A clearly identified ferrugsonite mode is the stretching $A_g$ mode arising from the scheelite $\nu_1(A_g)$ mode (F18 in **Table III**). This high-frequency stretching mode is experimentally found around 940 cm$^{-1}$ between 7.5 GPa and 8.2 GPa and is the responsible for the apparent sudden broadening of the scheelite $\nu_1(A_g)$ Raman peak at these two pressures. Therefore, the mode near 940 cm$^{-1}$ is indeed a wide double mode composed of a low-frequency $A_g$ mode of the ferrugsonite phase and a high-frequency $\nu_1(A_g)$ mode of the scheelite phase. The instability of both phases at 9 GPa, according to *ab initio* calculations **[17]**, is reflected in the experimental spectra by the gradual disappearance of both peaks between 7.5 and 9 GPa [see **Fig. 5(b)**]. Additional support for the assignment of the high-frequency stretching $A_g$ mode of the ferrugsonite phase derived from the scheelite $\nu_1(A_g)$ mode comes on the one hand from the calculated frequency (935 cm$^{-1}$ at 8.2 GPa), and on the other hand from the comparison of the frequencies of the scheelite and ferrugsonite modes in BaWO$_4$ with those measured in CaWO$_4$ and SrWO$_4$ under high pressure, as discussed below.

A comparative study of the frequencies of the scheelite and ferrugsonite Raman modes in BaWO$_4$ and the corresponding modes in CaWO$_4$ and SrWO$_4$ **[11,12]** allows us to exploit the trends induced by cation substitution in the series Ca, Sr, Ba. In particular, the high-frequency stretching modes in the ferrugsonite structure seem to behave in a coherent way in the three compounds after the scheelite-to-ferrugsonite phase transition. The $A_g$ ferrugsonite mode arising from the $\nu_1(A_g)$ mode is harder than its predecessor in CaWO$_4$ **[11]**, has the same frequency at the phase transition than its predecessor in SrWO$_4$ **[12]**, and should be slightly below its predecessor in BaWO$_4$, as it has been indeed found and discussed above.

The opposite behavior is found for the ferguosonite high-frequency stretching A$_g$ mode arising from the scheelite $\nu_3$(B$_g$) mode. In CaWO$_4$, the A$_g$ mode is below the $\nu_3$(B$_g$) mode at the phase transition; in SrWO$_4$ the A$_g$ mode is at the same frequency that the $\nu_3$(B$_g$) mode at the phase transition; and in BaWO$_4$ it is expected to be above the scheelite $\nu_3$(B$_g$) mode at the phase transition. Since the scheelite $\nu_3$(B$_g$) mode in BaWO$_4$ at 7.5 GPa is at 844 cm$^{-1}$ (see arrows in **Fig. 5(b)**), it is reasonable to attribute the mode around 858 cm$^{-1}$ (see asterisk in **Fig. 5(b)**) to the ferguosonite A$_g$ mode derived from the scheelite $\nu_3$(B$_g$) mode. In this case, our calculated frequency for this mode is 25 cm$^{-1}$ (3%) below the experimental frequency.

As regards to the ferguosonite stretching B$_g$ modes arising from the doubly degenerate scheelite $\nu_3$(E$_g$) mode, it should be noted that this mode shows a strange behavior in CaWO$_4$ and SrWO$_4$ since only one of the two modes has been observed in these compounds, and the observed mode shows a nonlinear behavior with pressure **[11,12]**. The measured ferguosonite B$_g$ frequencies in CaWO$_4$ and SrWO$_4$ are below that of the $\nu_3$(E$_g$) mode at the phase transition. Instead, we have tentatively located the ferguosonite B$_g$ modes near 826 and 839 cm$^{-1}$ at 7.5 GPa, i.e., slightly above the scheelite $\nu_3$(E$_g$) mode at 818 cm$^{-1}$ at 7.5 GPa. Support for this assignment comes from the closeness between these two modes according to our calculations that again seem to underestimate the frequency by about 20-25 cm$^{-1}$ with respect to our experimentally measured values.

Other modes of the ferguosonite structure are the A$_g$ modes arising from the scheelite $\nu_2$(A$_g$) and $\nu_2$(B$_g$) modes around 332 cm$^{-1}$ at ambient pressure. In BaWO$_4$ these two scheelite modes are almost degenerate like in SrWO$_4$ **[12]**. However, the high intensity of the low-frequency member of this pair in the ferguson ite phase of CaWO$_4$ and SrWO$_4$ **[11,12]**, and their splitting in SrWO$_4$, in contrast with CaWO$_4$, where both modes are observed at the same frequency that their parent modes at the phase transition, lead us to think that the strong mode at 338 cm$^{-1}$ at 7.5 GPa [see asterisk in **Fig. 5(a)**] could be the low-frequency A$_g$ mode of this pair. This assignment is supported by our calculations that locate this mode at almost the same frequency as experimentally observed (see **Table III**). Regarding the high-frequency A$_g$ mode of this pair, it should be 20 cm$^{-1}$ above the lower one in BaWO$_4$ if one follows the increase of separation between the two A$_g$ modes with increasing the mass of the A cation in the series Ca, Sr, and Ba. Our *ab initio* calculations locate several A$_g$ modes some 10-20 cm$^{-1}$ above the lower one, but the small intensity of this mode and the overlapping with other stronger modes assigned to the BaWO$_4$-II phase prevents us from observing it. The disappearance of the A$_g$ mode of 338 cm$^{-1}$ at 8.2 GPa has precluded us to measure its frequency pressure coefficient.

With respect to the three ferguson ite modes coming from the scheelite $\nu_4$(B$_g$+E$_g$) modes, they have been found between 450 and 500 cm$^{-1}$ in CaWO$_4$ and SrWO$_4$ **[11,12]**, therefore we thought that they should be also at similar frequencies in BaWO$_4$. However, our *ab initio* calculations suggest that these three modes are close to the two modes arising from the $\nu_2$(A$_g$) and $\nu_2$(B$_g$) modes below 400 cm$^{-1}$. The smaller distance between the scheelite $\nu_2$ and $\nu_4$ modes in BaWO$_4$ as compared to CaWO$_4$ and SrWO$_4$ supports the closeness of their related ferguson ite modes as obtained in our calculations. Likewise to the $\nu_2$ modes, the overlapping of the $\nu_4$ modes with other modes of higher intensity, assigned to the BaWO$_4$-II phase, has precluded us to measure their behavior under pressure.

As regards to the fergusonite modes related to the external modes in the scheelite phase, one fergusonite mode with high intensity in SrWO$_4$ is the A$_g$ mode arising from the external scheelite R(A$_g$) mode **[12]**. This fergusonite A$_g$ mode is at a lower frequency than its parent mode at the phase transition in CaWO$_4$ and at the same frequency in SrWO$_4$ so we expect it at slightly higher frequency than its parent in BaWO$_4$. On this basis, we think that this mode could be the 192 cm$^{-1}$ mode at 7.5 GPa. Furthermore, the increase of the pressure coefficient of this mode when increasing the A cation mass from 2 cm$^{-1}$/GPa in CaWO$_4$ to 3.3 cm$^{-1}$/GPa in BaWO$_4$ and the location of this mode at 181 cm$^{-1}$ at 8.2 GPa according to our *ab initio* calculations support the present assignment. Curiously enough, our calculations locate the two fergusonite B$_g$ modes related to the external scheelite R(E$_g$) mode around 233 cm$^{-1}$; however, we have not observed evidence of their presence. This result is similar to what has been reported in CaWO$_4$ and SrWO$_4$, where these two fergusonite modes were not found **[11,12]**. At present we have no idea why these two modes have not been observed.

Another fergusonite mode with high intensity in SrWO$_4$ is the A$_g$ mode arising from the topmost scheelite T(B$_g$) mode **[12]**. This A$_g$ mode is expected to be at slightly lower frequency in BaWO$_4$ than its parent at the phase transition. Since the T(B$_g$) mode is at 160.6 cm$^{-1}$ at 6.9 GPa, we think that the related A$_g$ mode could be a weak mode near 161 cm$^{-1}$ at 7.5 GPa. Support for this assignment comes from the location of this mode at 158 cm$^{-1}$ at 8.2 GPa according to our calculations. The remaining fergusonite A$_g$ mode in BaWO$_4$ would be the one arising from the lowest scheelite T(B$_g$). We have found a fergusonite mode with negative pressure coefficient at 37 cm$^{-1}$ at 7.5 GPa. We believe this is the lowest fergusonite A$_g$ mode, despite our calculations locate it between 38 and 56 cm$^{-1}$ depending on the approximation to the exchange-correlation potential used, LDA or GGA-PBE, respectively. It is rather striking that this mode with negative pressure coefficient has not been observed neither in CaWO$_4$ nor in SrWO$_4$ where the mode derived from the lowest scheelite T($B_g$) mode has positive pressure coefficient [11,12]. We think that the observation of this fergusonite mode with negative pressure coefficient in BaWO$_4$ is indicative of the instability of the fergusonite phase at high pressures in BaWO$_4$ as compared to CaWO$_4$ and SrWO$_4$. This result is in agreement with the smaller stability of the fergusonite phase at high pressures, as compared to the BaWO$_4$-II phase, already obtained from previous calculations [17].

Finally, as regards to the fergusonite $B_g$ modes arising from the two scheelite T($E_g$) modes, we have observed two close modes at 57 and 69 cm$^{-1}$ at 7.5 GPa that could correspond to the fergusonite $B_g$ modes coming from the lowest T($E_g$) mode. The mode at 57 cm$^{-1}$ appears at 7.5 GPa just above the frequency of the lowest scheelite T($B_g$) mode with negative Grüneisen parameter. The mode at 69 cm$^{-1}$ is observed as a shoulder of a double peak corresponding to the two high-pressure phases. Our calculations for these two modes are 83 and 84 cm$^{-1}$ at 8.2 GPa. Support for this assignment comes from the behavior of these two modes in the series Ca, Sr, Ba. In CaWO$_4$ the two $B_g$ modes derived from the topmost scheelite T($E_g$) mode are one well above and the other slightly below the frequency of the parent mode at the phase transition [11]. In SrWO$_4$ only one mode, that we believe is the mode with higher frequency, is found slightly above the frequency of the parent mode at the phase transition [12]. Therefore, we expect both fergusonite $B_g$ modes being below the frequency of the T($E_g$) mode at the phase transition in BaWO$_4$. As regards to the two fergusonite $B_g$ modes coming from the topmost scheelite T($E_g$) mode, we have found two modes that appear as shoulders near 93 and 118 cm$^{-1}$ at 7.5 GPa and that we have attributed to the fergusonite phase. Our calculations locate these two modes around 120 cm$^{-1}$ at 8.2 GPa. Our assignment for these latter modes can not be supported by the Raman measurements in CaWO$_4$ and SrWO$_4$ because only one mode out of two was observed in CaWO$_4$ [11], while none of the two modes was observed in SrWO$_4$ [12].

**B.2. BaWO$_4$-II phase**

The possibility for a high-pressure structure in BaWO$_4$ having the BaWO$_4$-II structure (SG No. 14) **[6]** has been recently suggested by Errandonea *et al*. **[17]**. The experimental powder ADXRD patterns above 11 GPa are compatible with the BaWO$_4$-II structure **[17]**. The appearance of Raman modes at 6.9 GPa; i.e. *before* the appearance of modes corresponding to the fergusonite phase at 7.5 GPa, is consistent with the *ab initio* total-energy calculations, which yield a lower value (5.1 GPa) for the scheelite/BaWO$_4$-II coexistence pressure than for the scheelite/fergusonite one (7.5 GPa) **[17]**. Therefore, we attribute the new Raman modes that appear at 6.9 GPa and last up to 16 GPa to the BaWO$_4$-II phase [marked by exclamations signs in **Figs. 5(a) and (b)**]. In this section, we will show with the help of our theoretical study that the Raman modes of BaWO$_4$ appearing above 6.9 GPa and remaining up to the highest pressures achieved of about 16 GPa are indeed compatible with the high-pressure phase BaWO$_4$-II. In particular, the BaWO$_4$-II phase shows a distorted octahedral W-O coordination which can give account for the Raman modes observed above 6.9 GPa in the phonon gap of the scheelite and fergusonite phases between 400 and 800 cm$^{-1}$. Similar modes have also been observed in the phase after the fergusonite structure taking place in BaMoO$_4$ above 9 GPa **[37,38]**.

The BaWO$_4$-II structure is centrosymmetric and has space group P2$_1$/n ($C_{2h}^5$) (SG No. 14) with Z = 8 **[6]**. The vibrational modes have the following mechanical representation at Γ **[26]**:

$$\Gamma = 36A_g + 36 A_u + 36B_g + 36B_u \qquad (6)$$

with 72 Raman-active (g) modes and 72 IR-active (u) modes, of which one A$_u$ and two B$_u$ are zero-frequency acoustic modes. One thus expects four times more Raman modes in the BaWO$_4$-II structure than in the fergusonite structure. The experimental assignment of the mode symmetry in the BaWO$_4$-II phase is difficult because of the lack of information on polarization inside the DAC, and because the number of modes that can be clearly resolved in the experimental Raman spectra above 6.9 GPa is around 40; i.e., about half the number of expected modes for the BaWO$_4$-II phase.

We think that the non-observation of all modes in the BaWO$_4$-II structure can be due to a number of factors, namely: 1) limited spectral resolution (since many peaks are very close to one another, especially those forming pairs, as evidenced by our *ab initio* calculations); 2) the small scattering cross section of some modes; and 3) overlapping with modes of the fergusonite phase and decrease in intensity of the BaWO$_4$-II modes above 9 GPa. In any case, the observation of at least three of the four *ab initio* predicted modes for the BaWO$_4$-II phase in certain wavenumber ranges, especially in the phonon gap of the scheelite and fergusonite phases, suggests that there is no other structure with higher symmetry, like monoclinic LaTaO$_4$ or raspite phases with SG No. 14 and with half the number of expected modes than the BaWO$_4$-II phase, which could give account for the experimentally observed modes.

**Figure 3** shows the pressure dependence of the Raman modes observed experimentally and attributed to the BaWO$_4$-II phase (black triangles). **Table V** summarizes the experimentally observed frequencies of the Raman modes attributed to the BaWO$_4$-II phase at 9 GPa. The assignment of the modes corresponding to the BaWO$_4$-II phase has been done with the help of *ab initio* lattice dynamics calculations of this phase at 6.9 GPa. **Tables VI** and **VII** summarize the calculated frequencies and symmetries of the Raman and IR modes attributed to the BaWO$_4$-II phase at 6.9 GPa, respectively. The calculated frequencies of the Raman modes in the BaWO$_4$-II phase at 6.9 GPa are also marked at the bottom of **Figs. 5(a) and (b)**.

As previously commented, the assignment of the high-pressure phase of BaWO$_4$ stable between 6.9 GPa and 16 GPa to the BaWO$_4$-II phase is supported by the observation, in several cases, of the four peaks expected for each fergusonite mode at frequencies close to those obtained from *ab initio* lattice dynamics calculations. Clear examples of this fact can be observed in the high-frequency region above 500 cm$^{-1}$, due to the smaller density of modes in this region. The Raman spectra of the high-pressure phase above 9 GPa show two strong high-frequency modes, located at 906 and 923 cm$^{-1}$ at 9 GPa, plus two weak high-frequency modes, located at 950 and 960 cm$^{-1}$ at 9 GPa. The two weak modes appear as shoulders rather than as peaks. In particular, the broad band around 950 cm$^{-1}$ is not well observed because it overlaps with the scheelite $\nu_1(A_g)$ and fergusonite $A_g$ modes, and it can only be observed above 9 GPa once the scheelite and fergusonite modes fade (see **Fig. 4**). *Ab initio* calculations for the BaWO$_4$-II phase at 6.9 GPa show that many modes in this phase are grouped in pairs [see marks at the bottom of **Figs. 5(a) and (b)**]. In particular, there are two groups of high-frequency modes located around 880 and 920 cm$^{-1}$ at 6.9 GPa approximately 25 cm$^{-1}$ (3%) below our experimentally observed modes. Therefore, we attribute these four high-frequency modes to the modes in the BaWO$_4$-II phase that likely arise from the splitting of the scheelite $\nu_1(A_g)$ mode or the fergusonite $A_g$ mode. This is the clearest example we have found for the observation of four modes in the BaWO$_4$-II phase for every fergusonite mode, what supports the BaWO$_4$-II nature of the high-pressure phase. In fact, *ab initio* calculations show four fergusonite modes and twelve BaWO$_4$-II modes above 600 cm$^{-1}$ [see **Fig. 5(b)**], what is in agreement with the 4:1 BaWO$_4$-II/fergusonite mode ratio, and with the assignment done above.

The two modes of the *ab initio* calculated pairs can be experimentally observed in some other cases above 500 cm$^{-1}$. For instance, there are two pairs of calculated modes around 625 cm$^{-1}$ at 6.9 GPa. Again, one can distinguish three of the four modes between 600 and 680 cm$^{-1}$ at pressures between 6.9 and 9.0 GPa [see exclamation marks in **Fig. 5(b)**] and a fourth one above 10 GPa. In a similar way, there is a pair of calculated modes slightly below 800 cm$^{-1}$ at 6.9 GPa. We attribute the two modes of the pair to the most prominent peaks slightly above 800 cm$^{-1}$, that dominate the Raman spectrum above 10 GPa (see **Fig. 4** and **Fig. 5(b)**). Furthermore, we attribute the weak mode located at 877 cm$^{-1}$ at 9 GPa and visible in all the spectra between 7.5 and 16 GPa (see **Table V**) to the single calculated mode located at 830 cm$^{-1}$ at 6.9 GPa; i.e., underestimated in calculations by almost 40 cm$^{-1}$ (5%). Similarly, there are two pairs of calculated modes around 500 cm$^{-1}$ at 6.9 GPa [see marks at the bottom of **Fig. 5(b)**]. One can distinguish three of the four modes between 470 and 530 cm$^{-1}$ at pressures between 6.9 and 9.0 GPa [see exclamation marks in **Fig. 5(b)**]. In addition, a fourth one can be also distinguished above 10 GPa. All these examples give evidence that, despite the number of Raman modes distinguished in the experimental high-pressure spectra is well below the 72 modes expected for the BaWO$_4$-II phase, this phase can be reasonably identified.

The presence of two strong and two weak high-frequency stretching modes above 10 GPa arising from the strong scheelite $\nu_1(A_g)$ mode, and located at 906 and 923 cm$^{-1}$, and at 950 and 960 cm$^{-1}$ at 9 GPa, respectively, suggests that the rule by which strong modes in a high-pressure phase derive from strong modes in the low pressure phase, already observed for many modes of the fergusonite phase seems not be valid for the BaWO$_4$-II phase derived from the scheelite phase. We think that the reason why the strong scheelite $\nu_1(A_g)$ mode splits into two strong and two weak modes in the BaWO$_4$-II structure is related to the change of the atomic symmetry in these two phases. As commented previously, the $A_g$ modes of the scheelite structure only have contribution from the vibration of the O modes in a 16f Wyckoff position. However, in the BaWO$_4$-II structure all atoms occupy a 4e Wyckoff position and contribute to the $A_g$ and $B_g$ modes [36]. Therefore, it is expected that the four modes derived from a strong mode in the scheelite structure do not have the same scattering cross section and some of them modes could have small Raman scattering cross sections. This could be one of the reasons why we observe approximately half the number of Raman modes that were expected for the BaWO$_4$-II phase.

A striking feature of the transformation to the BaWO$_4$-II phase can be clearly seen in the Raman spectra of 7.5 and 8.2 GPa near 330 cm$^{-1}$. The fergusonite mode located at 338 cm$^{-1}$ at 7.5 GPa disappears at 8.2 GPa. Instead, a mode around 330 cm$^{-1}$, not present before, appears in the spectrum at 8.2 GPa. The difference in frequency of these modes on going from the fergusonite to the BaWO$_4$-II structure is in complete agreement with our lattice dynamics calculations that show a fergusonite mode around 339 cm$^{-1}$ at 8.2 GPa that leads to four modes with the lowest one at 327 cm$^{-1}$ in the BaWO$_4$-II phase at 6.9 GPa [see **Fig. 5(a)**].

We should mention that the coexistence of the three phases: scheelite, fergusonite, and BaWO$_4$-II is possible due to the kinetic hindrance of the reconstructive scheelite-to-BaWO$_4$-II phase transition and the displacive second-order nature of the scheelite-to-fergusonite phase transition [39]. The kinetic hindrance of the scheelite-to-BaWO$_4$-II transition can be clearly observed in our Raman spectra between 6.9 and 10 GPa by checking the intensity of the two strong high-frequency peaks attributed to the BaWO$_4$-II phase. It can be observed that the low-frequency mode already appears at 6.9 GPa, followed by the high-frequency mode at 7.5 GPa. However, both modes do not achieve their final intensity up to 10 GPa. Therefore, we can conclude that the transformation to the BaWO$_4$-II structure is completed above 9.5-10 GPa (see dashed line in **Fig. 3**). This result is in agreement with ADXRD measurements that report the x-ray diffraction patterns indexed with the BaWO$_4$-II structure above 11 GPa [17]. Furthermore, we think that the BaWO$_4$–II structure is retained up to 16 GPa (maximum pressure attained in this study) and that the decrease in signal with increasing pressure can be due to a tendency of the sample to amorphisation before the completion of the next phase transition, as already suggested in **Ref. 17** on the basis of the loss of intensity of the x-ray pattern observed above 10 GPa.

In summary, we finish the B section by concluding that the onset of the scheelite-to-BaWO$_4$-II phase transition is found around 6.9 GPa, being followed by a scheelite-to-fergusonite phase transition around 7.5 GPa. These phase transition pressures are in excellent agreement with the pressure of the scheelite-to-BaWO$_4$-II transition (5.1 GPa) and the pressure of the scheelite-to-fergusonite transition (7.5 GPa) found recently by *ab initio* total-energy calculations **[17]**. They also agree with the observation of a first phase transition at 6.8 GPa in previous Raman scattering measurements **[7]**, and of two phase transitions around 7 and 11 GPa in powder ADXRD measurements **[14,17]**. The observation of the coexistence of the scheelite, fergusonite, and BaWO$_4$-II structures between 7.5 GPa and 9.0 GPa can only be explained by the hindrance of the reconstructive scheelite-to-BaWO$_4$-II phase transition (due to its slow kinetics likely caused by an activation barrier) that favours the observation of the second-order scheelite-to-fergusonite phase transition, as already discussed in **Ref. 17 and 39**.

**C. Tungsten coordination in high-pressure phases**

Our assignment of the Raman modes in the high-pressure phases to the fergusonite and BaWO$_4$-II phases is coherent with the change of W coordination from tetrahedral to octahedral with increasing pressure. The feature that suggests a tetrahedral W coordination in the fergusonite phase and an octahedral W coordination in the BaWO$_4$–II phase is the appearance of modes in the phonon gap of the scheelite and fergusonite structures between 400 and 800 cm$^{-1}$ after the scheelite-to-BaWO$_4$-II phase transition. Another feature supporting this change in coordination is the decrease of the strong highest-frequency stretching $A_g$ mode from 945 cm$^{-1}$, in the scheelite phase, to 940 cm$^{-1}$, in the ferrusonite phase, and to 900 and 920 cm$^{-1}$ in the BaWO$_4$–II phase at the phase transition. The small decrease of the high-frequency $A_g$ stretching mode in the scheelite-to-ferrusonite transition means that there is a small increase of the shorter W-O bond distance after the phase transition that is still compatible with the tetrahedral coordination of W ions in the ferrusonite structure. However, the large decrease of the highest-frequency modes (more than 20 cm$^{-1}$) in the scheelite-to-BaWO$_4$-II transition suggests a large increase of the shortest W-O bond distances that must be due to the change of the coordination of W ions from tetrahedral to octahedral. This was also evidenced by XANES measurements above 9.8 GPa **[17]**.

According to Hardcastle and Wachs, there is a relationship between the frequencies of the stretching W-O modes and the bond distance R (in Å) between W and O in tungsten oxides **[40]**:

$$\omega \text{ (cm}^{-1}\text{)} = 25823 \exp(-1.902 \cdot R) \tag{7}$$

On the other hand, according to Brown and Wu, there is a relationship between the bond distance R (in Å) and the Pauling's bond strengths, which for tungsten oxides is **[41]**:

$$s_{W-O} = (R/1.904)^{-6} \tag{8}$$

with $s_{W-O}$ given in valence units (v.u.). It is possible to get an empirical relationship between the Raman stretching mode frequency and the Pauling's W-O bond strengths using **Eqs. (7) and (8)**:

$$s_{W-O} = [0.27613 \ln(25823/\omega)]^{-6} \tag{9}$$

thus, following Hardcastle and Wachs, it is possible to estimate the coordination of the W ion in a tungsten oxide material if we know all the stretching frequencies of the material, that usually fall in the high-frequency region above 400 cm$^{-1}$ **[40]**.

By taking as the stretching frequencies of the scheelite structure those of the modes at 795, 831, and 926 cm$^{-1}$ at 1 atm (see **Table I**), we have estimated W-O bonding distances of 1.83, 1.81, and 1.75 Å, giving bond strengths of 1.27, 1.37, and 1.66 v.u., respectively. With these numbers and knowing that the W coordination in the scheelite structure is fourfold we get the estimated total valence of 1.27 + 1.37 + 1.66 + 1.66 = 5.96 $\cong$ 6 that is the formal valence of the W ion. To get this result we have had to consider a double contribution of the shorter W-O bond distance (1.75 Å). In such a case, we get an average W-O bond distance of 1.784 Å that agrees with the estimated W-O bond distance from XRD measurements, and with the expected ideal W-O bond distance in tetrahedral coordination (1.78 Å), which corresponds to a Pauling's bond strength of 1.5 v.u. in **Eq. (8)** for each of the four W-O bonds **[41]**.

Similarly, by taking as the stretching frequencies of the fergusonite structure the modes at 826, 839, 859 and 940 cm$^{-1}$ at 7.5 GPa (see **Table III**), we have estimated W-O bond distances of 1.81, 1.80, 1.79 and 1.74 Å giving strengths of 1.36, 1.39, 1.45 and 1.71 v.u., respectively. With these numbers we have estimated a total valence of 5.91 for W, that is very close to the formal valence of 6, and evidences that the distinction between external and internal modes in the fergusonite phase could be still applicable at least in BaWO$_4$ **[40]**, and that the configuration for the W ions is a tetrahedral configuration, as expected in the fergusonite structure.

Since the fergusonite structure retains the tetrahedral coordination of the W ion, we can use the same but reverse reasoning to CaWO$_4$ and SrWO$_4$ in order to estimate the frequency of the lost fergusonite high-frequency B$_g$ stretching mode. For this purpose we have taken the frequencies of 800, 865, and 950 cm$^{-1}$ as those of the stretching modes in fergusonite CaWO$_4$ at 15 GPa. With these data and **Eq. (9)** one would obtain a total valence of 6 if an additional stretching mode was located around 878 cm$^{-1}$ at 15 GPa. In the case of fergusonite SrWO$_4$, we have taken the frequencies of 790, 860, and 950 as those of the stretching modes in fergusonite SrWO$_4$ at 15 GPa. Again, one would obtain a total valence of 6 if an additional stretching mode was located around 890 cm$^{-1}$ at 15 GPa. In CaWO$_4$, the proximity of the predicted mode at 878 cm$^{-1}$ and the mode found at 865 cm$^{-1}$ lead us to think that this last mode could be a double mode indeed. A close inspection of the spectra of **Ref. 11** suggests a broadening of the mode at 865 cm$^{-1}$ and the possible presence of a mode around 860 cm$^{-1}$ at 15 GPa. If this is indeed a double mode, our estimated result would be in extreme good agreement with the observation given the small accuracy of the frequency (around 5%) that can be estimated with **Eq. (9) [40]**. In SrWO$_4$, we can not say anything because of the limited definition of the reported experimental spectra **[12]**. In any case, the tetrahedral W coordination in the three alkaline-earth fergusonite tungstates CaWO$_4$, SrWO$_4$, and BaWO$_4$ is guaranteed by the shorter average W-O distances found with Hardcastle and Wachs' formula **[40]**. According to this formula and the available experimental Raman modes, the average W-O bond distances for fergusonite CaWO$_4$, SrWO$_4$, and BaWO$_4$ at 15, 15, and 7.5 GPa are 1.783, 1.785, and 1.786 Å, respectively. These average W-O distances agree with the ideal W-O distance for tetrahedral coordination (1.78 Å) **[41]**.

The application of Hardcastle and Wachs' rules to the BaWO$_4$-II can help in assigning the first-order stretching modes of this low symmetry structure if the WO$_6$ octahedra can be regarded as almost independent units **[40]**. In the characterization of the BaWO$_4$-II structure at 1 atm, the following W-O bond distances were reported: 1.83, 1.83, 1.84, 1.84, 1.86, 1.88, 1.96, 1.97, 2.07, 2.13, 2.18, and 2.33 Å corresponding to two non-equivalent WO$_6$ octahedra **[6]**. The BaWO$_4$-II structure shows a 4+2 coordination for W with four W-O distances between 1.8 and 2.0 Å, and two longer distances between 2.1 and 2.3 Å. These distances give an average W-O distance of 1.977 Å, which is quite deviated from the ideal W-O bond distance for octahedral W coordination (1.904 Å), and quite close to the ideal W-O bond distance for eightfold W coordination (1.99 Å) given by **Eq. (8)**. If we take the six average W-O bond distances for the BaWO$_4$-II structure at 1 atm given in **Ref. 6** we obtain: 1.83, 1.84, 1.87, 1.965, 2.10, and 2.255 Å, and application of **Eq. (8)** yields the following average Pauling's bond strengths: 1.27, 1.23, 1.11, 0.83, 0.55, and 0.36 v.u. Altogether they sum 5.35 v.u., which do not fully agree with the 6 valence of the W ion **[40]**. Furthermore, **Eq. (7)** allows us to calculate the stretching frequencies yielding: 795, 780, 751, 723, 621, 609, 504, 449, 408, and 307 cm$^{-1}$. However, these values do not agree neither with the experimentally observed high-frequency Raman modes above 6.9 GPa nor with the calculated Raman modes at 6.9 GPa by first principles (see **Table V**). The deviation is significant for the modes with higher frequencies that are experimentally found around 900 cm$^{-1}$, even considering the difference in pressure between 1 atm **[6]** and our observations and calculations. The good behaviour of **Eqs. (7) and (8)** in the case of the scheelite and fergusonite structures, and the knowledge that the W coordination of the BaWO$_4$-II phase must be sixfold, lead us to think that the previous estimations for the BaWO$_4$-II phase fail due to two possible reasons: 1) the W-O bond distances reported in **Ref. 6**, in particular the shortest ones, could be somewhat overestimated; 2) **Eq. (7)** is not valid for the BaWO$_4$-II structure because the WO$_6$ octahedra can not be regarded as independent units and consequently the distinction between external and internal modes is no longer valid in this complex structure, and one can not deal with pure stretching modes.

Since we do not know other experimental structural parameters of the BaWO$_4$-II phase, we only can try to get a deeper insight of the BaWO$_4$-II phase if we calculate the Raman stretching frequencies and bond strengths with **Eqs. (7) and (8)** considering the structural parameters at 9 GPa obtained by *ab initio* total-energy calculations **[17]**. With those data we get the following W-O bond distances: 1.78, 1.80, 1.80, 1.81, 1.85, 1.88, 1.93, 1.95, 2.13, 2.14, 2.15, and 2.22 Å corresponding to the two non-equivalent $WO_6$ octahedra. These distances give an average W-O distance of 1.954 Å, which is intermediate between the ideal W-O bond distance for octahedral W coordination (1.904 Å) and the ideal W-O bond distance for eightfold W coordination (1.99 Å) **[41]**. If we take the six average W-O bond distances for the $BaWO_4$-II structure at 9 GPa we obtain: 1.79, 1.81 1.86, 1.94, 2.14, and 2.18 Å, and application of **Eq. (8)** yields the following average Pauling's bond strengths: 1.45, 1.39, 1.14, 0.90, 0.50, and 0.44 v.u. Altogether they sum 5.82 v.u., which agree with the 6 valence of the W ion. Furthermore, **Eq. (7)** allows us to calculate the stretching frequencies yielding: 878, 840, 821, 773, 723, 661, 629, 609, 446, 442, 436, and 380 $cm^{-1}$. These values are relatively close to experimentally measured values of Raman modes of the $BaWO_4$-II. In this sense, it must be noted that **Eq. (7)** tends to underestimate the frequency of the Raman modes, especially in the high-frequency region. According to **Eq. (7)** the Raman frequency corresponding to the stretching mode of the ideal $WO_4$ tetrahedra is 874 $cm^{-1}$. This is close to the *average* stretching frequency in the scheelite phase (870 $cm^{-1}$) and in the ferguson ite phase (866 $cm^{-1}$) of $BaWO_4$, but it is far from the frequency of the highest stretching mode in the scheelite and ferguson ite phases (around 940 $cm^{-1}$). Therefore, one has to deal carefully with estimations given by Eqs. (7) and (8). With this consideration in mind, we can conclude that our above estimated frequencies obtained from the structural data of the $BaWO_4$-II phase according to *ab initio* total-energy calculations at 9 GPa, correlate relatively well with experimentally reported values if we add typically some 20 to 40 $cm^{-1}$ to the estimated frequencies. Finally, we can conclude that structural data of the $BaWO_4$-II phase at 9 GPa according to *ab initio* total-energy calculations **[17]** give a rather accurate description of the $BaWO_4$-II structure that allowed us to verify the validity of **Eqs. (7) and (8)**. Furthermore, we consider that either the W-O bond distances reported in **Ref. 6**, in particular the shortest ones, are somewhat overestimated, or they correspond to a W coordination higher than six for this structure at ambient conditions.

**D. Phase transitions in other related compounds**

In order to analyze the high-pressure phases of BaWO$_4$ in comparison with other related compounds, we want to mention that we have shown in the B.1 section that our results for scheelite-BaWO$_4$ are consistent with those previously reported by Jayaraman *et al.* **[7]**. Here we will show that our results for the fergusonite and BaWO$_4$-II phases can also give account for the Raman spectra of BaWO$_4$ at 6.8 and 8.4 GPa reported by Jayaraman *et al.* **[7]**. To us, the Raman spectrum at 6.8 GPa in **Ref. 7** corresponds to a mixture of the three phases. At 6.8 GPa, the scheelite phase is the dominant one, but evidence of the BaWO$_4$-II phase is given by the 901 cm$^{-1}$ peak. Evidence of the fergusonite phase at this pressure is given by the broadening of the 944.5 cm$^{-1}$ peak, as compared to the same peak at 5.6 GPa, and by the presence of new peaks near the scheelite ones, like the mode that we observed at 338 cm$^{-1}$ at 7.5 GPa. As regards to the Raman spectrum at 8.4 GPa in **Ref. 7**, it mainly corresponds to the BaWO$_4$-II phase, since the high-frequency modes of the scheelite and fergusonite phases have completely disappeared, and also the mode at 338 cm$^{-1}$ of the fergusonite phase is also gone. However, it is strange for us that they do not observe the strong high-frequency mode of the BaWO$_4$-II phase near 920 cm$^{-1}$. In fact, one can guess it from the high-frequency shoulder of the 906 cm$^{-1}$ peak at 8.4 GPa, but it is not clearly present in the spectra of **Ref. 7**. Support for the assignment of the spectrum at 8.4 GPa in **Ref. 7** to the BaWO$_4$–II phase comes from the appearance of new modes near 302, 330, and 403 cm$^{-1}$ similar to ours. In summary, we think that the Raman spectra of BaWO$_4$ in **Ref. 7**, agree qualitatively with ours with only two differences: 1) our Raman spectra show larger resolution than those of Jayaraman *et al*. what explains our observation of more modes than in their spectra; and 2) the phase transitions we have seen in their work appear at slightly smaller pressures than those reported in this work.

The fergusonite nature of the Raman spectra of $CaWO_4$ and $SrWO_4$ above 12 GPa and of $BaWO_4$ above 7.5 GPa is also partially supported by the comparison with the Raman spectrum of $HgWO_4$ at 1 atm [42]. $HgWO_4$ is a compound crystallizing in a monoclinic structure belonging to space group No. 15. In $HgWO_4$, the W ion has octahedral coordination, but it can be regarded as tetrahedral since two W-O distances are far more large than the other four [42]. In $HgWO_4$, there is a high-frequency stretching mode around 930 $cm^{-1}$; i.e., at a similar frequency than the high-frequency fergusonite $A_g$ mode found in the three alkaline-earth tungstates. There are also other two high-frequency modes at 850 and 815 $cm^{-1}$. These two modes agree with the frequencies observed in $CaWO_4$ and $SrWO_4$ and are about 10 $cm^{-1}$ below two of the modes we have found in $BaWO_4$ at 7.5 GPa. The main difference in the stretching modes between the four compounds is found in the lowest-frequency mode that is located near 700 $cm^{-1}$ in $HgWO_4$, i.e., at much smaller frequency than those found in the alkaline-earth tungstates. It is possible that the 700 $cm^{-1}$ mode is likely due to the tendency to octahedral coordination in tungstates with high electronegative A cations like Hg and Pb, which tends to close the phonon gap of the scheelite and fergusonite phases. As regards to the low-frequency modes in $HgWO_4$, there are a couple of modes near 500 $cm^{-1}$ that could be part of the three fergusonite modes arising from the scheelite $\nu_4(B_g+E_g)$ modes, already found at similar frequencies in $CaWO_4$ and $SrWO_4$ but expected at smaller frequencies in $BaWO_4$. In $HgWO_4$, there are also modes observed in the 200-400 $cm^{-1}$ region that could be related to those in the same region in alkaline earth-tungstates. The modes at 380 and 335 $cm^{-1}$ are likely to be the two fergusonite modes coming from the scheelite $\nu_2(A_g+B_g)$ mode, that are located at similar frequencies in CaWO$_4$ and SrWO$_4$ at 15 GPa. The two strong modes at 285 and 300 cm$^{-1}$ will likely correspond to the two fergusonite modes derived from the scheelite R(E$_g$) mode that have not been observed in the alkaline-earth tungstates but are expected in that region. The mode at 235 cm$^{-1}$ is likely to be the fergusonite mode derived from the scheelite R(A$_g$) mode, and the strong mode at 180 cm$^{-1}$ in HgWO$_4$ is likely to be an A$_g$ mode deriving from the topmost scheelite T(B$_g$) mode. This last assignment is due to the tendency in AWO$_4$ scheelite tungstates with heavy A cations to have the R(A$_g$) mode above the topmost T(B$_g$) mode **[29]**. In any case, new and more detailed RT Raman spectra of monoclinic fergusonite-like HgWO$_4$ would be of help in the definite assignment of all its modes, especially in the low frequency region.

Finally, it is also interesting to compare the behavior of the Raman modes in BaWO$_4$ under pressure with those reported for BaMoO$_4$ under pressure. Several works have recently appeared reporting the Raman spectra of BaMoO$_4$ under pressure **[37,38,43]**. In a similar fashion to the recent powder ADXRD studies of BaWO$_4$ under pressure, recent powder ADXRD measurements in BaMoO$_4$ under pressure have shown a first phase transition to the fergusonite structure around 6 GPa **[37]**. Such a phase transition is also characterized by the appearance of new Raman modes near the frequency of the scheelite $\nu_1$(A$_g$) mode, the splitting of the scheelite $\nu_3$ modes, and the appearance of many modes between 300 and 400 cm$^{-1}$, and also near 100 and 200 cm$^{-1}$ **[37,38,43]**. It can be observed that almost all modes in the new phase form groups of bands. Furthermore, as in BaWO$_4$, a second phase transition has been observed in BaMoO$_4$ above 9 GPa **[37,43]**. The second high-pressure phase has been suggested to be a monoclinic structure with octahedral W-O coordination and space group P2$_1$/c (SG No. 14). The Raman spectrum of the second high-pressure phase is characterized by several strong bands near the frequency of the scheelite $\nu_1$(A$_g$) mode, the increase of the intensity of a couple of bands near the frequency of the scheelite $\nu_3$ modes, the appearance of several broad bands between 600 and 700 cm$^{-1}$, and the appearance of a single mode near 440 cm$^{-1}$. These features look like similar, at similar frequencies, and at similar pressures than those observed by us in the Raman spectra of BaWO$_4$ above 6.9 GPa. Therefore, considering that the pressure behavior of BaMoO$_4$ and BaWO$_4$ should be similar on the light of their similar positions in Fukunaga and Yamaoka's [44] and in Bastide's [45] diagrams, we believe that a close inspection of the Raman spectra of BaMoO$_4$ at pressures between 6 and 10.8 GPa could lead to the observation of the ferguson­ite and the BaWO$_4$-II type structures as we have found in this work.

**V. Conclusions**

We have performed RT Raman scattering measurements under pressure in BaWO$_4$ up to 16 GPa. The frequency pressure dependence of all first-order modes of the scheelite phase have been measured up to the onset of the scheelite-to-BaWO$_4$-II phase transition around 6.9 GPa. This value of the transition pressure is in good agreement with the calculated value of the I/II coexistence pressure (5.1 GPa) according to recent *ab initio* total-energy calculations [17]. Our measurements show that the transition to the BaWO$_4$-II phase is not completed up to 9.5 GPa. Besides, we have found a phase transition towards the metastable fergusonite phase at 7.5 GPa in complete agreement with *ab initio* total-energy calculations (7.5 GPa) [17]. In summary, we have observed the following structural sequences in BaWO$_4$: 1) scheelite between 1 atm and 6.9 GPa; 2) scheelite+BaWO$_4$-II (incomplete transition) between 6.9 and 7.5 GPa, 3) scheelite + BaWO$_4$-II + fergusonite between 7.5 and 9.0 GPa, and finally 4) BaWO$_4$-II from 9.5 GPa up to 16 GPa. On decreasing pressure from 16 GPa, we have found that the scheelite structure is recovered below 3 GPa in agreement with ADXRD measurements [17].

The observation of the scheelite-to-fergusonite phase transition at pressures above the scheelite-to-BaWO$_4$-II phase transition can only be explained by the kinetic hindrance of the reconstructive scheelite-to-BaWO$_4$-II phase transition due to its slow kinetics likely caused by an activation barrier, and the displacive second-order nature of the scheelite-to-fergusonite phase transition, as already pointed out in **Ref. 17** and discussed in **Ref. 39**. As a consequence of this, we have observed the coexistence of the scheelite, fergusonite and BaWO$_4$-II phases in the pressure range of 7.5 to 9.0 GPa. The phase coexistence observed in the present Raman study, but not observed in previous ADXRD measurements, may be related to: 1) the higher sensitivity of Raman measurements to local phases with respect to ADXRD; and 2) the use of "large" single crystals of BaWO$_4$ in our Raman measurements. The presence of dislocation glides is a typical feature of scheelite-structured orthotungstates **[21]**. The dislocation density is expected to increase and propagate along "large" single crystals under an applied stress, eventually favouring the transformation of different parts of the scheelite sample into different metastable polymorphs. However, the increase and propagation of dislocations is unlikely to take place in micron-size powder samples which would explain the lack of phase coexistence in previous ADXRD experiments **[17]**. Therefore, the previous x-ray diffraction results of Errandonea *et al.* and Panchal *et al.* can be understood on the light of the present study. The Raman peaks of the BaWO$_4$-II phase appear before those of the fergusonite, but once the fergusonite phase appears its Raman peaks are stronger than those of the BaWO$_4$-II phase and dominate the spectrum. This result explains why in the ADXRD study **[17]** the fergusonite phase was only observed within the 7.5-9 GPa pressure interval of phase coexistence and why the ADXRD pattern of the BaWO$_4$-II phase appears only after the fergusonite phase extinguishes above 9 GPa.

Finally, we have performed *ab initio* lattice dynamics calculations of BaWO$_4$ at selected pressures in the scheelite, fergusonite, and BaWO$_4$-II phases. Our calculated mode frequencies in the three structures agree with the frequencies of the observed Raman modes and have allowed the assignment and discussion of the nature of many modes in the three phases. The Raman spectrum of the fergusonite phase in $BaWO_4$ compare well with those of the fergusonite phase in $CaWO_4$ and $SrWO_4$. The intrinsic dynamic instability of the fergusonite phase in $BaWO_4$ is supported by the presence of an external soft mode at an even lower frequency than that of the scheelite phase, and by the rapid decrease of the intensity of the $A_g$ mode derived from the scheelite $\nu_1(A_g)$ mode with increasing pressure. Furthermore, we have shown with the help of Hardcastle and Wachs', and of Brown and Wu's formulae, that the fergusonite phase of alkaline-earth tungstates has tetrahedral W coordination while the $BaWO_4$-II phase has an octahedral coordination for the W cation, and that the $WO_4$ tetrahedra in the fergusonite phase and the $WO_6$ octahedra in the $BaWO_4$-II phase can be regarded as almost independent units.

**Acknowledgments**

The authors thank Dr. P. Bohacek (Institute of Physics, Prague) for providing $BaWO_4$ crystals. This work was made possible through financial support of the MCYT of Spain under grants number: MAT2004-05867-C03-03, MAT2004-05867-C03-01, and MAT2002-04539-C02-02. F.J.M. acknowledges financial support by the "*Programa Incentivo a la Investigación de la U.P.V.*" D.E. and N.G. acknowledge the financial support from the MCYT of Spain through the "Ramon y Cajal" program. The use of the computational resources of the Barcelona Supercomputer Center (Mare Nostrum) for the DFT calculations is also gratefully acknowledged. J.L.S., A.M., and P. R-H. acknowledge the financial support from the Consejería de Educación del Gobierno Autónomo de Canarias.

# References

[1] M. Bravin, M. Bruckmayer, C. Bucci, S. Cooper, S. Giordano, F. von Feilitzsch, J. Hohne, J. Jochum, V. Jorgens, R. Keeling, H. Kraus, M. Loidl, J. Lush, J. Macallister, J. Marchese, O. Meier, P. Meunier, U. Nagel, T. Nussle, F. Probst, Y. Ramachers, H. Sarsa, J. Schnagl, W. Seidel, I. Sergeyev, M. Sisti, L. Stodolsky, S. Uchaikin, and L. Zerle, Astroparticle Physics **12**, 107 (1999).

[2] G. Angloher, C. Bucci, C. Cozzini, F. von Feilitzsch, T. Frank, D. Hauff, S. Henry, T. Jagemann, J. Jochum, H. Kraus, B. Majorovits, J. Ninkovic, F. Petricca, F. Probst, Y. Ramachers, W. Rau, W. Seidel, M. Stark, S. Uchaikin, L. Stodolsky, and H. Wulandari, Nucl. Instrum. Methods Phys. Res. A **520**, 108 (2004).

[3] A. Alessandrello, V. Bashkirov, C. Brofferio, C. Bucci, D. V. Camin, O. Cremonesi, E. Fiorini, G. Gervasio, A. Giuliani, A. Nucciotti, M. Pavan, G. Pessina, E. Previtali, L. Zanotti, Phys. Lett. B **420,** 109 (1998).

[4] C. Cozzini, G. Angloher, C. Bucci, F. von Feilitzsch, D. Hauff, S. Henry, T. Jagemann, J. Jochum, H. Kraus, B. Majorovits, V. Mikhailik, J. Ninkovic, F. Petricca, W. Potzel, F. Probst, Y. Ramachers, W. Rau, M. Razeti, W. Seidel, M. Stark, L. Stodolsky, A. J. B. Tolhurst, W. Westphal, H. Wulandari, Phys. Rev. C **70**, 064606 (2004).

[5] A.W. Sleight, Acta Cryst. B **28**, 2899 (1972).

[6] I. Kawada, K. Kato, and T. Fujita, Acta Crist. B **30**, 2069 (1974).

[7] A. Jayaraman, B. Batlogg, and L.G. Van Uitert, Phys. Rev. B **28**, 4774 (1983).

[8] A. Jayaraman, B. Batlogg, and L.G. Van Uitert, Phys. Rev. B **31**, 5423 (1985).

[9] M. Nicol, and J. F. Durana, J. Chem. Phys. **54**, 1436 (1971).

[10] B.N. Ganguly and M. Nicol, phys. stat. sol. (b) **79**, 617 (1977).

[11] D.Christofilos, S. Ves, and G.A. Kourouklis, phys. stat. sol. (b) **198**, 539 (1996).

[12] D.Christofilos, K. Papagelis, S. Ves, G.A. Kourouklis, and C. Raptis, J. Phys.: Condens. Matter **14**, 12641 (2002).

[13] A. Grzechnik, W.A. Crichton, M. Hanfland, and S. Van Smaalen, J. Phys.: Condens. Matter **15**, 7261 (2003).

[14] V. Panchal, N. Garg, A.K. Chauhan, Sangeeta, and S. M. Sharma, Solid State Commun. **130**, 203 (2004).

[15] A. Grzechnik, W.A. Crichton, and M. Hanfland, phys. stat. sol. (b) **242**, 2795 (2005).

[16] D. Errandonea, J. Pellicer-Porres, F.J. Manjón, A. Segura, Ch. Ferrer, R.S. Kumar, O. Tschauner, P. Rodríguez-Hernández, J. López-Solano, S. Radescu, A. Mujica, A. Muñoz, and G. Aquilanti, Phys. Rev. B **72**, 174106 (2005).

[17] D. Errandonea, J. Pellicer-Porres, F.J. Manjón, A. Segura, Ch. Ferrer, R.S. Kumar, O. Tschauner, P. Rodríguez-Hernández, J. López-Solano, S. Radescu, A. Mujica, A. Muñoz, and G. Aquilanti, Phys. Rev. B **73**, (2006).

[18] D. Errandonea, F.J. Manjón, M. Somayazulu, and D. Häusermann, J. Solid State Chem. **177**, 1087 (2004).

[19] J. M. Besson, J. P. Itié, A. Polian, and G. Weill, J. L. Mansot, J. Gonzalez, Phys. Rev. B **44**, 4214 (1991).

[20] M. Nikl, P. Bohacek, N. Mihokova, M. Kobayashi, M. Ishii, Y. Usuki, V. Babin, A. Stolovich, S. Zazubovich, and M. Bacci, J. Lumin. **87-89**, 1136 (2000).

[21] P. Mogilevsky, Phylosophical Magazine **85**, 3511 (2005).

[22] H.K. Mao, J. Xu, and P.M. Bell, J. Geophys. Res. **91**, 4673 (1986).

[23] G. Kresse et al., computer code VASP. For more information see: http://cms.mpi.univie.ac.at/vasp.

[24] G. Kresse and J. Joubert, Phys. Rev. B **59**, 1758 (1999).

[25] D. Alfe, G.D. Price, and M.J. Gillan, Phys. Rev. B **64**, 045123 (2001).

[26] D.L. Rousseau, R.P. Baumann, and S.P.S. Porto, J. Raman Spect. **10**, 253 (1981).

[27] S.P. S. Porto, and J.F. Scott, Phys. Rev. **157**, 716 (1967).

[28] G. Herzberg, *Molecular Spectra and Molecular Structure II: Infra-Red and Raman Spectra* (D. Van Nostrand Co. Inc., New York, 1945).

[29] M. Liegegois-Duyckaerts and P. Tarte, Spectrochim. Acta **28A**, 2037 (1972).

[30] A.N. Akimov, M.V. Nikanovich, V.G. Popov, and D.S. Umreiko, J. Appl. Spectr. **45**, 806 (1987).

[31] N. Weinstock, H. Schulze, and A. Müller, J. Chem. Phys. **59**, 5063 (1973).

[32] F.J. Manjón, D. Errandonea, A. Segura, V. Muñoz, G. Tobías, P. Ordejón, and E. Canadell, Phys. Rev. B **63**, 125330 (2001).

[33] P.J. Miller, R.K. Khanna, and E.R. Lippincott, J. Phys. Chem. Solids 34, 533 (1973).

[34] P. Tarte and M. Liegeois-Duyckaerts, Spectr. Acta **28 A**, 2029 (1972).

[35] F.M Pontes, M.A.M.A. Maurera, A.G. Souza, E. Longo, E.R. Leite, R. Magnani, M.A.C. Machado, P.S. Pizani, and J.A. Varela, J. Euro. Ceram. Soc. **23**, 3001 (2003).

[36] Spectral Active Modes in Bilbao Crystallographic Server. http://www.cryst.ehu.es

[37] V. Panchal, N. Garg, and S. M. Sharma, J. Phys.: Condens. Matter **18**, 3917 (2006).

[38] D.Christofilos, J. Arvanitidis, E. Kampasakali, K. Papagelis, S. Ves, and G.A. Kourouklis, phys. stat. sol. (b) **241**, 3155 (2004).

[39] D. Errandonea, submitted to J.Phys. Chem. Solids.

[40] F.D. Hardcastle and I.E. Wachs, J. Raman Spectr. **26**, 397 (1995).

[41] I.D. Brown and K.K. Wu, Acta Crystallogr. Sect. B **32**, 1957 (1976).

[42] G. Blasse, J. Inorg. Nucl. Chem **37**, 97 (1975).

[43] D.Christofilos, J. Arvanitidis, E. Kampasakali, K. Papagelis, S. Ves, and G.A. Kourouklis, Proceedings of the Joint 20th AIRAPT – 43th EHPRG Meeting (2005).

[44] O. Fukunaga and S. Yamaoka, Phys. Chem. Minerals **5**, 167 (1979).

[45] J.P. Bastide, J. Solid State Chem. **71**, 115 (1987).

**Table I.** *Ab initio* calculated and experimental zero-pressure frequencies, pressure coefficients, and Grüneisen parameters of the Raman modes in scheelite BaWO$_4$. For obtaining the experimental Grüneisen parameter, $\gamma = B_0/\omega \cdot d\omega/dP$, we have taken the zero-pressure bulk modulus of scheelite BaWO$_4$, $B_0 = 52$ GPa **[17]**. Raman modes in other scheelite alkaline-earth tungstates SrWO$_4$ and CaWO$_4$ are also given for comparison.

| Peak/mode | BaWO$_4$ (theory) | | | BaWO$_4$ (experiment) | | | SrWO$_4$ | | | CaWO$_4$ | | |
|---|---|---|---|---|---|---|---|---|---|---|---|---|
| | $\omega(0)$ cm$^{-1}$ | $d\omega/dP$ cm$^{-1}$/GPa | $\gamma$ | $\omega(0)$ cm$^{-1}$ | $d\omega/dP$ cm$^{-1}$/GPa | $\gamma$ | $\omega(0)$ cm$^{-1}$ | $d\omega/dP$ cm$^{-1}$/GPa | $\gamma$ | $\omega(0)$ cm$^{-1}$ | $d\omega/dP$ cm$^{-1}$/GPa | $\gamma$ |
| T(B$_g$) | 55 | -1.4 | -1.40 | 63 | -0.8 | -0.67[a], -0.64[b] | 75 | -0.4[c] | -0.66[c], -0.32[b] | 84 | -0.4[d] | -0.43[d], -1.2[b] |
| T(E$_g$) | 81 | 1.0 | 0.63 | 74 | 1.0 | 0.73[a], 0.43[b] | 102 | 1.3[c] | 0.78[c], 1.00[b] | 116 | 1.7[d] | 1.33[d], 1.10[b] |
| T(E$_g$) | 110 | 3.2 | 1.42 | 101 | 3.3 | 1.70[a], 0[b]* | 135 | 2.9[c] | 1.23[c], 1.73[b] | 196 | 3.7[d] | 1.72[d], 2.10[b] |
| T(B$_g$) | 145 | 4.1 | 1.50 | 133 | 4.1 | 1.58[a] | 171 | 3.4[c] | 1.23[c] | 227 | 4.7[d] | 1.88[d] |
| R(A$_g$) | 149 | 5.4 | 1.70 | 150 | 4.2 | 1.47[a], 1.17[b] | 190 | 4.4[c] | 1.42[c], 1.55[b] | 212 | 3.8[d] | 1.63[d], 1.70[b] |
| R(E$_g$) | 209 | 6.9 | 1.64 | 191 | 6.3 | 1.71[a] | 238 | 6.8[c] | 1.76[c], 1.70[b] | 276 | 7.0[d] | 2.31[d] |
| $\nu_2$(A$_g$) | 328 | 3.0 | 0.46 | 331 | 2.5 | 0.40[a], 0.38[b] | 337 | 3.3[c] | 0.60[c], 0.58[b] | 334 | 2.5[d] | 0.68[d], 0.65[b] |
| $\nu_2$(B$_g$) | 329 | 3.9 | 0.56 | 332 | 3.0 | 0.46[a], 0.38[b] | 337 | 3.3[c] | 0.60[c], 0.58[b] | 334 | 2.5[d] | 0.68[d], 0.65[b] |
| $\nu_4$(B$_g$) | 339 | 2.7 | 0.44 | 344 | 2.0 | 0.30[a], 0.23[b] | 371 | 4.1[c] | 0.68[c], 0.62[b] | 402 | 4.1[d] | 0.93[d], 1.10[b] |
| $\nu_4$(E$_g$) | 348 | 3.8 | 0.57 | 352 | 3.4 | 0.50[a] | 380 | 4.6[c] | 0.74[c], 0.64[b] | 406 | 4.6[d] | 1.03[d] |
| $\nu_3$(E$_g$) | 797 | 3.0 | 0.20 | 795 | 3.2 | 0.21[a], 0.16[b] | 799 | 3.0[c] | 0.23[c], 0.25[b] | 797 | 3.0[d] | 0.34[d], 0.34[b] |
| $\nu_3$(B$_g$) | 823 | 2.0 | 0.13 | 831 | 2.0 | 0.12[a], 0.11[b] | 837 | 2.1[c] | 0.15[c], 0.14[b] | 838 | 1.9[d] | 0.21[d], 0.17[b] |
| $\nu_1$(A$_g$) | 928 | 2.8 | 0.16 | 926 | 2.7 | 0.15[a], 0.12[b] | 921 | 2.2[c] | 0.15[c], 0.16[b] | 911 | 1.5[d] | 0.15[d], 0.14[b] |

[a] This work, [b] **Ref. 7** after correcting the bulk modulus of BaWO$_4$, [c] **Ref.11**, [d] **Ref. 12**, *See text.

**Table II.** Mode frequencies, pressure coefficients, and Grüneisen parameters of the calculated IR modes in scheelite-BaWO$_4$ at zero-pressure. Experimentally measured IR modes (TO,LO) at RT are also reported for comparison. The frequencies of the silent B$_u$ modes are also shown for completeness.

| Peak/mode | ω(0) (cm$^{-1}$) | dω/dP (cm$^{-1}$/GPa) | γ | ω(0) (exp.) (cm$^{-1}$) |
|---|---|---|---|---|
| T(A$_u$) | 0 | - | - | 0 |
| T(E$_u$) | 0 | - | - | 0 |
| T(E$_u$) | 101 | -1.0 | -0.51 | (98,104)$^a$, 109$^b$ |
| T(A$_u$) | 128 | 3.6 | 1.46 | (110,113)$^a$, 129$^b$ |
| R(E$_u$) | 141 | 3.8 | 1.40 | (135,159)$^a$, 142$^b$ |
| R(B$_u$) | 196 | 0.4 | 0.11 | |
| ν$_4$(A$_u$) | 250 | 2.0 | 0.42 | (258,296)$^a$, 283$^b$ |
| ν$_4$(E$_u$) | 297 | 3.0 | 0.53 | (292,317)$^a$, 313$^b$ |
| ν$_2$(A$_u$) | 375 | 4.6 | 0.64 | (374,384)$^a$, 379$^b$ |
| ν$_2$(B$_u$) | 387 | 5.5 | 0.74 | |
| ν$_3$(A$_u$) | 801 | 2.7 | 0.18 | (790,886)$^a$, 828$^b$, 887$^c$ |
| ν$_3$(E$_u$) | 807 | 2.8 | 0.18 | (796,894)$^a$, 828$^b$, 806$^c$ |
| ν$_1$(B$_u$) | 934 | 2.4 | 0.13 | |

$^a$**Ref. 33** (single crystal), $^b$**Ref. 34** (powder samples), $^c$**Ref. 35** (powder samples).

**Table III.** Frequencies at 7.5 GPa and zero-pressure coefficients of the Raman modes of fergusonite-BaWO$_4$ as obtained from fittings to the data using $\omega(P) = \omega(0) + d\omega/dP \cdot P$. The relative frequency pressure coefficients at zero pressure are also given for comparison. The fergusonite frequencies obtained after *ab initio* calculations at 8.2 GPa are also given for comparison.

| Peak/mode | $\omega(7.5)$ cm$^{-1}$ | $d\omega/dP$ cm$^{-1}$/GPa | $1/\omega \cdot d\omega/dP$ | $\omega(8.2)^a$ cm$^{-1}$ | $d\omega/dP$ $^a$ cm$^{-1}$/GPa |
|---|---|---|---|---|---|
| F1(A$_g$) | 37.5(4) | -0.58(3) | -0.014 | 56 | -1.8 |
| F2(B$_g$) | 59(1) | 0.83(6) | 0.016 | 83 | 1.2 |
| F3(B$_g$) | 67(1) | 0.58(4) | 0.009 | 84 | 0.6 |
| F4(B$_g$) | 93(1) | 2.22(8) | 0.029 | 121 | 2.4 |
| F5(B$_g$) | 118(1) | 1.23(5) | 0.011 | 122 | 2.6 |
| F6(A$_g$) | 161(1) | 1.75(5) | 0.011 | 158 | 2.9 |
| F7(A$_g$) | 192(1) | 3.33(9) | 0.020 | 181 | 5.6 |
| F8(B$_g$) | | | | 233 | 6.2 |
| F9(B$_g$) | | | | 234 | 6.4 |
| F10(A$_g$) | 338(1) | - | - | 339 | 2.5 |
| F11(A$_g$) | | | | 342 | 2.7 |
| F12(A$_g$) | | | | 352 | 3.8 |
| F13(B$_g$) | | | | 362 | 3.9 |
| F14(B$_g$) | | | | 363 | 3.9 |
| F15(B$_g$) | 826(2) | 1.79(5) | 0.002 | 809 | 2.3 |
| F16(B$_g$) | 839(3) | 4.09(9) | 0.005 | 810 | 3.0 |
| F17(A$_g$) | 859(1) | 0.50(2) | 0.0006 | 833 | 1.2 |
| F18(A$_g$) | 940(1) | 0.52(1) | 0.0006 | 935 | 0.1 |

$^a$ *Ab initio* calculations

**Table IV.** IR mode frequencies and pressure coefficients in fergusonite-BaWO$_4$ as obtained from *ab initio* calculations at 8.2 GPa.

| Peak/mode | ω(8.2) cm$^{-1}$ | dω/dP cm$^{-1}$/GPa |
|---|---|---|
| F1(A$_u$) | 0 | - |
| F2(B$_u$) | 0 | - |
| F3(B$_u$) | 0 | - |
| F4(A$_u$) | 99 | 0.6 |
| F5(B$_u$) | 101 | 0.05 |
| F6(B$_u$) | 138 | 2.2 |
| F7(A$_u$) | 155 | 2.3 |
| F8(B$_u$) | 156 | 2.4 |
| F9(B$_u$) | 204 | 0.6 |
| F10(A$_u$) | 248 | 5.3 |
| F11(A$_u$) | 309 | 3.1 |
| F12(A$_u$) | 310 | 3.2 |
| F13(B$_u$) | 394 | 4.0 |
| F14(B$_u$) | 409 | 4.9 |
| F15(B$_u$) | 812 | 2.1 |
| F16(B$_u$) | 817 | 2.3 |
| F17(A$_u$) | 818 | 2.1 |
| F18(A$_u$) | 933 | 2.5 |

**Table V.** Frequencies at 9 GPa and zero-pressure coefficients of the Raman modes observed in the BaWO$_4$-II phase as obtained from fittings to the data using ω(P) = ω(0)+ dω/dP·P. The relative frequency pressure coefficients at zero pressure are also given for comparison.

| Peak /mode | ω(9) (cm$^{-1}$) | Dω/dP (cm$^{-1}$/GPa) | 1/ω·dω/dP | Peak /mode | ω(9) (cm$^{-1}$) | dω/dP (cm$^{-1}$/GPa) | 1/ω·dω/dP |
|---|---|---|---|---|---|---|---|
| B1 | 70.6(7) | 0.3(3) | 0.004 | B22 | 435(3) | 4.3(5) | 0.009 |
| B2 | 98.5(6) | 0.5(8) | 0.005 | B23 | 478(3) | 2.9(5) | 0.006 |
| B3 | 130.3(5) | 1.2(9) | 0.010 | B24 | 495(2) | 4.0(9) | 0.008 |
| B4 | 152(2) | 2.1(7) | 0.016 | B25 | 532(2) | 2.3(2) | 0.004 |
| B5 | 164(2) | 2.6(4) | 0.019 | B26 | 546(3) | 4.3(7) | 0.010 |
| B6 | 181(2) | 2.6(4) | 0.016 | B27 | 614(3) | 2.1(3) | 0.004 |
| B7 | 209(3) | 2.5(6) | 0.013 | B28 | 634(3) | 2.1(4) | 0.003 |
| B8 | 218(2) | 1.7(1) | 0.008 | B29 | 658(2) | 1.5(2) | 0.002 |
| B9 | 234(2) | 2.7(6) | 0.013 | B30 | 673(2) | 0.5(4) | 0.0008 |
| B10 | 258(3) | 3.1(2) | 0.013 | B31 | 697(2) | 2.6(7) | 0.004 |
| B11 | 287(3) | 1.7(6) | 0.006 | B32 | 745(2) | 2.4(4) | 0.003 |
| B12 | 302(2) | 2.6(4) | 0.009 | B33 | 770(2) | 3.5(5) | 0.005 |
| B13 | 304(2) | 0.5(2) | 0.002 | B34 | 798(2) | 3.5(7) | 0.005 |
| B14 | 326(1) | 2.0(5) | 0.006 | B35 | 815(2) | 3.9(8) | 0.005 |
| B15 | 330(1) | 2.8(4) | 0.009 | B36 | 843(2) | 2.4(5) | 0.003 |
| B16 | 359(2) | 2.5(3) | 0.008 | B37 | 877(2) | 2.8(6) | 0.003 |
| B17 | 369(3) | 2.1(4) | 0.006 | B38 | 906(1) | 3.2(2) | 0.004 |
| B18 | 400(1) | 1.7(6) | 0.004 | B39 | 923(1) | 2.8(2) | 0.003 |
| B19 | 404(2) | 1.8(7) | 0.004 | B40 | 950(2) | 3.2(8) | 0.003 |
| B20 | 417(3) | 2.6(9) | 0.006 | B41 | 963(2) | 2.9(7) | 0.003 |
| B21 | 432(3) | 3.2(5) | 0.007 | | | | |

**Table VI.** Raman mode symmetries and frequencies in the BaWO$_4$-II phase as obtained from *ab initio* calculations at 6.9 GPa.

| Mode (sym) | ω(6.9) (cm$^{-1}$) | Mode (sym) | ω(6.9) (cm$^{-1}$) | Mode (sym) | ω(6.9) (cm$^{-1}$) | Mode (sym) | ω(6.9) (cm$^{-1}$) |
|---|---|---|---|---|---|---|---|
| R1(A$_g$) | 54 | R19(A$_g$) | 145 | R37(A$_g$) | 286 | R55(A$_g$) | 517 |
| R2(B$_g$) | 59 | R20(B$_g$) | 146 | R38(B$_g$) | 288 | R56(B$_g$) | 520 |
| R3(A$_g$) | 66 | R21(B$_g$) | 159 | R39(A$_g$) | 328 | R57(A$_g$) | 623 |
| R4(B$_g$) | 69 | R22(A$_g$) | 163 | R40(B$_g$) | 329 | R58(B$_g$) | 624 |
| R5(A$_g$) | 86 | R23(B$_g$) | 178 | R41(A$_g$) | 338 | R59(A$_g$) | 633 |
| R6(A$_g$) | 87 | R24(A$_g$) | 179 | R42(B$_g$) | 339 | R60(B$_g$) | 635 |
| R7(A$_g$) | 89 | R25(A$_g$) | 186 | R43(A$_g$) | 341 | R61(B$_g$) | 698 |
| R8(B$_g$) | 90 | R26(B$_g$) | 188 | R44(B$_g$) | 352 | R62(A$_g$) | 699 |
| R9(B$_g$) | 95 | R27(B$_g$) | 198 | R45(B$_g$) | 380 | R63(B$_g$) | 734 |
| R10(B$_g$) | 98 | R28(A$_g$) | 205 | R46(A$_g$) | 381 | R64(A$_g$) | 737 |
| R11(A$_g$) | 104 | R29(A$_g$) | 229 | R47(A$_g$) | 395 | R65(A$_g$) | 755 |
| R12(A$_g$) | 115 | R30(B$_g$) | 235 | R48(B$_g$) | 397 | R66(B$_g$) | 789 |
| R13(B$_g$) | 118 | R31(B$_g$) | 245 | R49(A$_g$) | 410 | R67(A$_g$) | 795 |
| R14(B$_g$) | 122 | R32(A$_g$) | 256 | R50(B$_g$) | 423 | R68(B$_g$) | 830 |
| R15(A$_g$) | 123 | R33(B$_g$) | 258 | R51(B$_g$) | 440 | R69(A$_g$) | 876 |
| R16(B$_g$) | 128 | R34(A$_g$) | 263 | R52(A$_g$) | 445 | R70(B$_g$) | 880 |
| R17(A$_g$) | 138 | R35(A$_g$) | 276 | R53(A$_g$) | 479 | R71(A$_g$) | 912 |
| R18(B$_g$) | 141 | R36(B$_g$) | 278 | R54(B$_g$) | 482 | R72(B$_g$) | 913 |

**Table VII.** IR mode symmetries and frequencies in the BaWO$_4$-II phase as obtained from *ab initio* calculations at 6.9 GPa.

| Mode (sym) | ω(6.9) (cm$^{-1}$) | Mode (sym) | ω(6.9) (cm$^{-1}$) | Mode (sym) | ω(6.9) (cm$^{-1}$) | Mode (sym) | ω(6.9) (cm$^{-1}$) |
|---|---|---|---|---|---|---|---|
| I1(B$_u$) | 0 | I19(B$_u$) | 133 | I37(B$_u$) | 299 | I55(B$_u$) | 447 |
| I2(A$_u$) | 0 | I20(A$_u$) | 134 | I38(A$_u$) | 302 | I56(A$_u$) | 461 |
| I3(B$_u$) | 0 | I21(A$_u$) | 149 | I39(A$_u$) | 308 | I57(A$_u$) | 586 |
| I4(B$_u$) | 42 | I22(B$_u$) | 160 | I40(B$_u$) | 311 | I58(B$_u$) | 588 |
| I5(A$_u$) | 46 | I23(A$_u$) | 161 | I41(B$_u$) | 338 | I59(B$_u$) | 595 |
| I6(A$_u$) | 64 | I24(B$_u$) | 164 | I42(A$_u$) | 341 | I60(A$_u$) | 600 |
| I7(B$_u$) | 73 | I25(A$_u$) | 185 | I43(B$_u$) | 356 | I61(B$_u$) | 716 |
| I8(A$_u$) | 75 | I26(B$_u$) | 187 | I44(A$_u$) | 358 | I62(A$_u$) | 719 |
| I9(B$_u$) | 81 | I27(B$_u$) | 202 | I45(A$_u$) | 381 | I63(B$_u$) | 735 |
| I10(B$_u$) | 84 | I28(A$_u$) | 203 | I46(B$_u$) | 386 | I64(A$_u$) | 748 |
| I11(A$_u$) | 91 | I29(B$_u$) | 217 | I47(B$_u$) | 392 | I65(B$_u$) | 774 |
| I12(A$_u$) | 93 | I30(A$_u$) | 219 | I48(A$_u$) | 394 | I66(A$_u$) | 788 |
| I13(B$_u$) | 99 | I31(A$_u$) | 236 | I49(A$_u$) | 406 | I67(A$_u$) | 798 |
| I14(A$_u$) | 105 | I32(B$_u$) | 240 | I50(B$_u$) | 407 | I68(B$_u$) | 828 |
| I15(B$_u$) | 110 | I33(A$_u$) | 257 | I51(B$_u$) | 412 | I69(B$_u$) | 866 |
| I16(A$_u$) | 113 | I34(B$_u$) | 260 | I52(A$_u$) | 414 | I70(A$_u$) | 870 |
| I17(A$_u$) | 121 | I35(B$_u$) | 285 | I53(B$_u$) | 434 | I71(B$_u$) | 894 |
| I18(B$_u$) | 123 | I36(A$_u$) | 288 | I54(A$_u$) | 444 | I72(A$_u$) | 908 |

# Figure captions

**Fig. 1** Room-temperature Raman spectra of the scheelite-type $BaWO_4$ at different pressures between 1 bar and 7.5 GPa. Arrows mark the new peaks appearing in the Raman spectrum above 6.9 GPa. Note that small interferences due to $N_2$ in air are observed in the all spectra below 150 cm$^{-1}$. The dashed line indicates the position of a plasma line of $Ar^+$ at 104 cm$^{-1}$ used for calibration of Raman spectra. The *ab initio* calculated frequencies of the scheelite Raman modes at 1 bar are marked at the bottom.

**Fig. 2** Detail of the scheelite $\nu_2$ mode in $BaWO_4$ at different pressures. The Raman spectra at 1.6, 4.1, 5.1, 6.2, and 6.9 GPa have been shifted by 2.6, 10.45, 13.4, 16.8, and 18.3 cm$^{-1}$, respectively in order to bring the stronger $\nu_2(B_g)$ mode into coincidence at all pressures. The inset shows the second derivative of the Raman spectra that have been used to follow the pressure dependence of the $\nu_2(A_g)$ and $\nu_4(B_g)$ modes with respect to the frequency of the $\nu_2(B_g)$ mode.

**Fig. 3** Pressure dependence of the Raman mode frequencies of the scheelite (circles), fergusonite (blank squares) and $BaWO_4$–II (triangles) phases of $BaWO_4$ up to 16 GPa. Dotted lines show the onset of the scheelite-to-$BaWO_4$-II phase transition and of the scheelite-to-fergusonite transition. Dashed line indicates the pressure for the completion of the scheelite-to-$BaWO_4$-II transition.

**Fig. 4** Room-temperature Raman spectra of the fergusonite and $BaWO_4$-II phases of $BaWO_4$ at different pressures between 6.9 and 16 GPa. Arrows indicate the position of some Raman peaks of the fergusonite phase that disappear above 9 GPa. Double arrows indicate the positions of some Raman peaks of the $BaWO_4$-II phase that appear above

6.9 GPa. The dashed line indicates the position of a plasma line of $Ar^+$ at 104 cm$^{-1}$ used for calibration of Raman spectra.

**Fig. 5** Detail of the Raman spectra of BaWO$_4$ between 6.9 and 9 GPa at low frequencies (a) and high frequencies (b). The dashed line indicates the position of a plasma line of $Ar^+$ at 104 cm$^{-1}$ used for calibration of Raman spectra. Arrows indicate the position of some weak modes of the scheelite phase. Asterisks indicate the position of the modes of the fergusonite phase. Exclamation marks indicate the position of peaks attributed to the BaWO$_4$–II phase. At the bottom of the figures the *ab initio* calculated frequencies of the fergusonite phase at 8.2 GPa and of the BaWO$_4$-II phase at 6.9 GPa are shown.

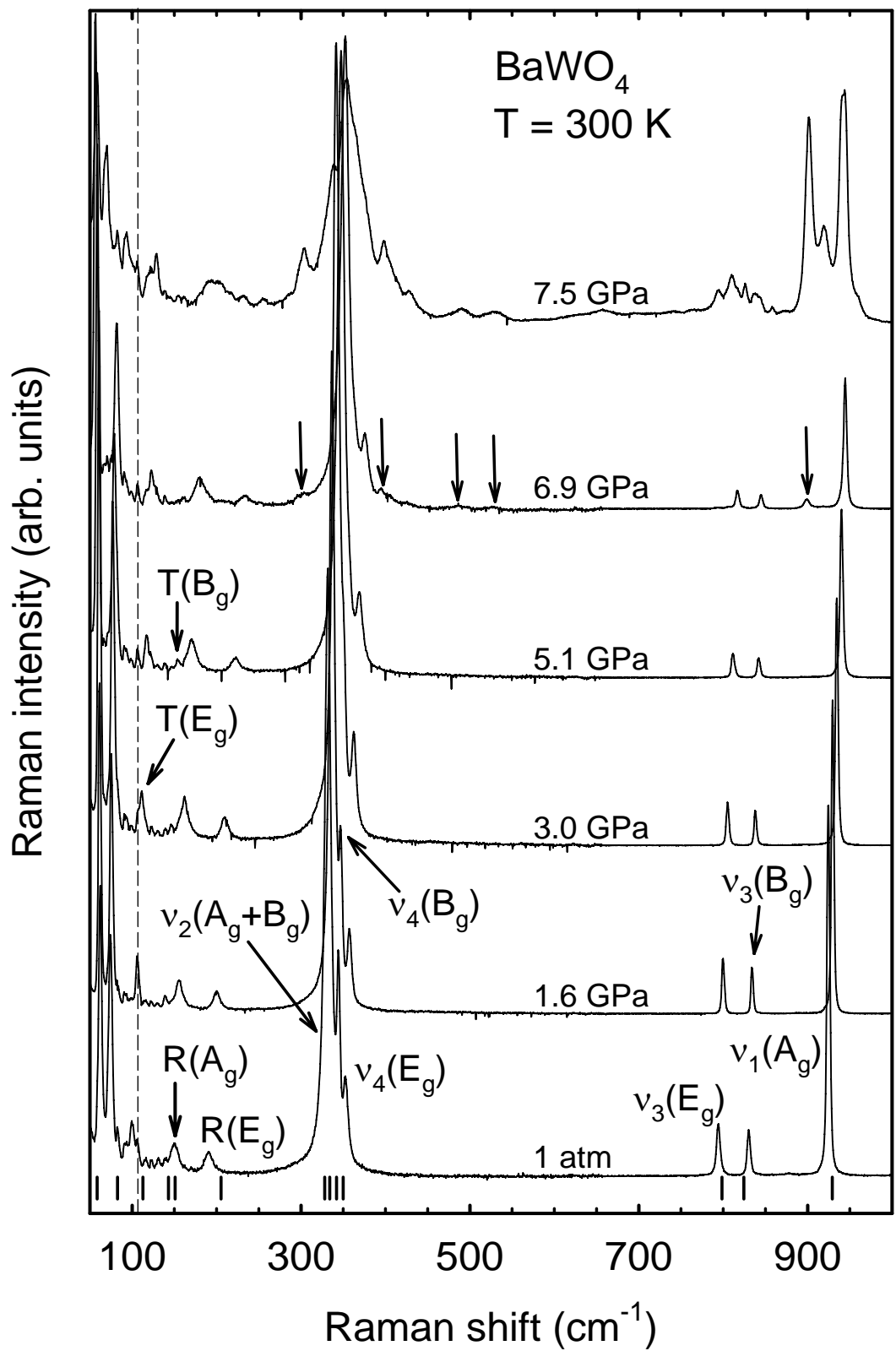

**Figure 1.** F.J. Manjón et al.

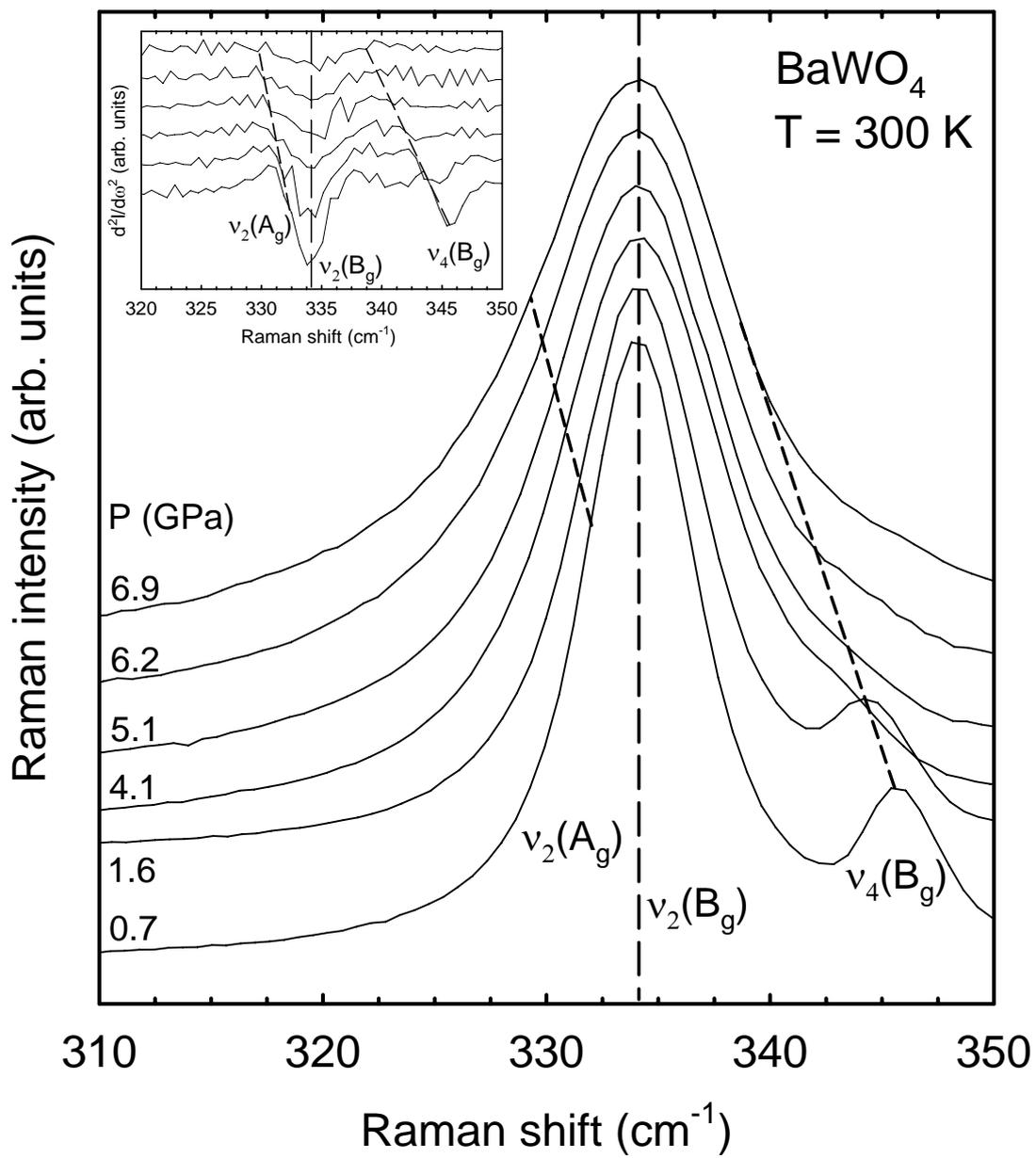

**Figure 2.** F.J. Manjón et al.

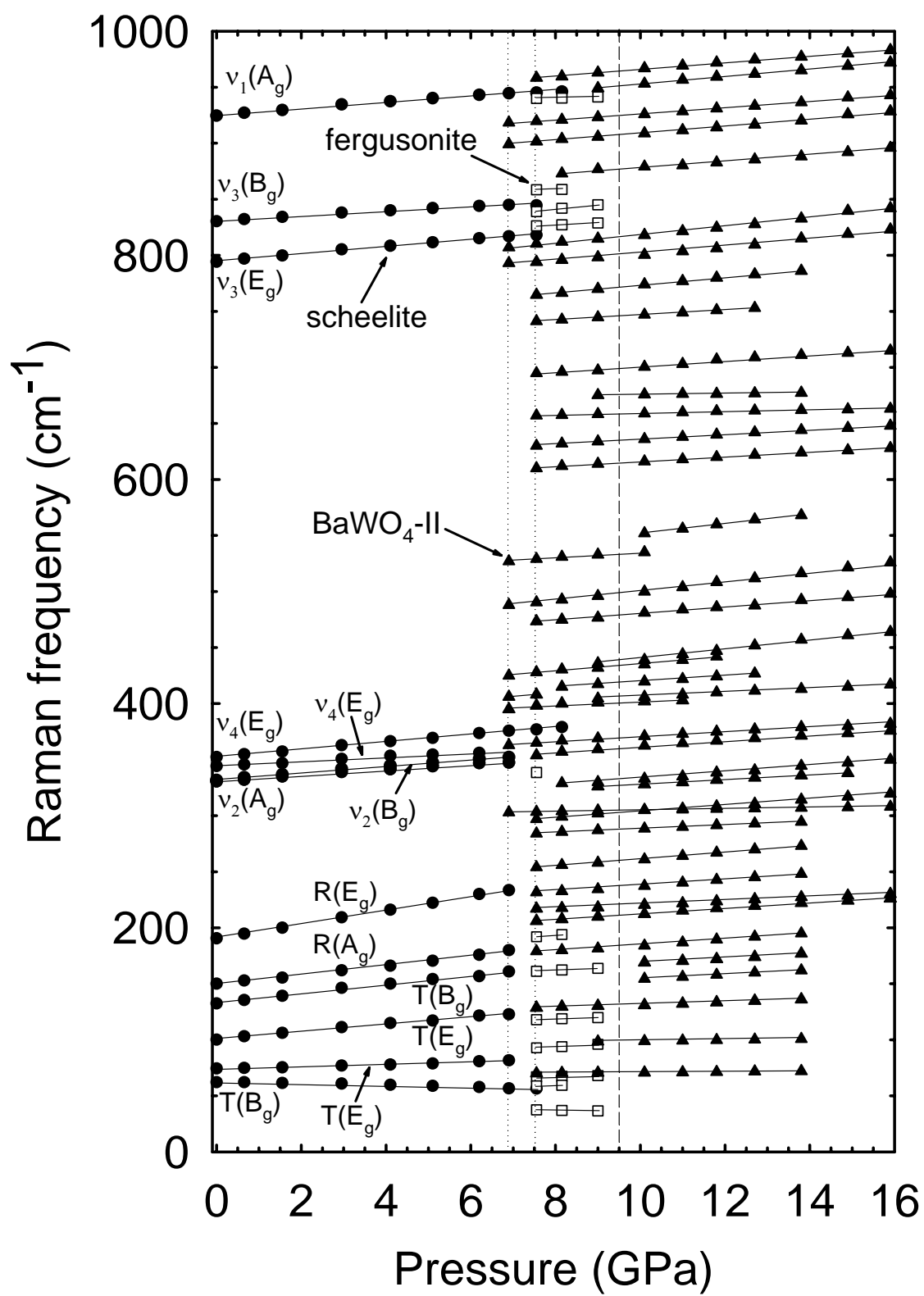

**Figure 3.** F.J. Manjón et al.

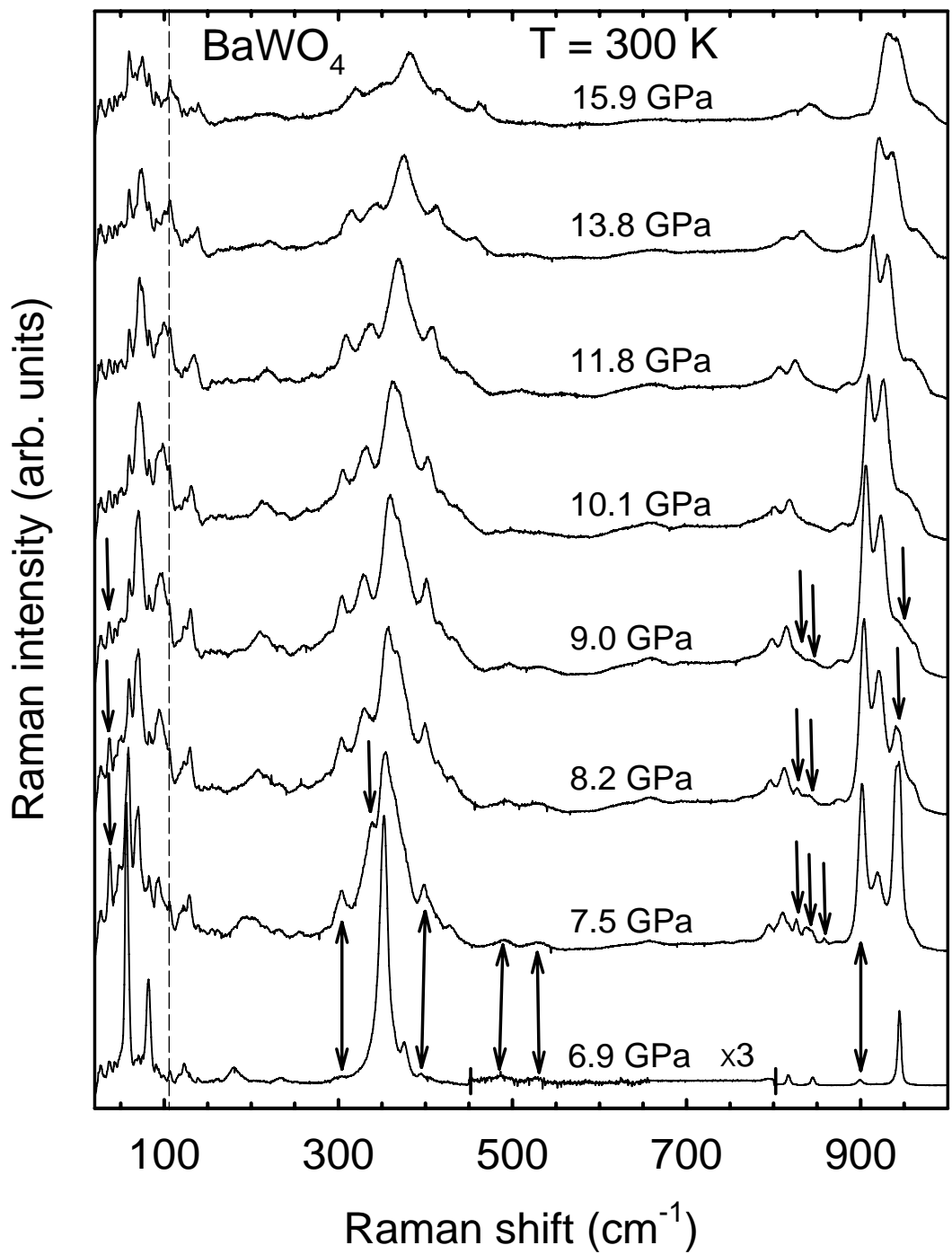

**Figure 4.** F.J. Manjón et al.

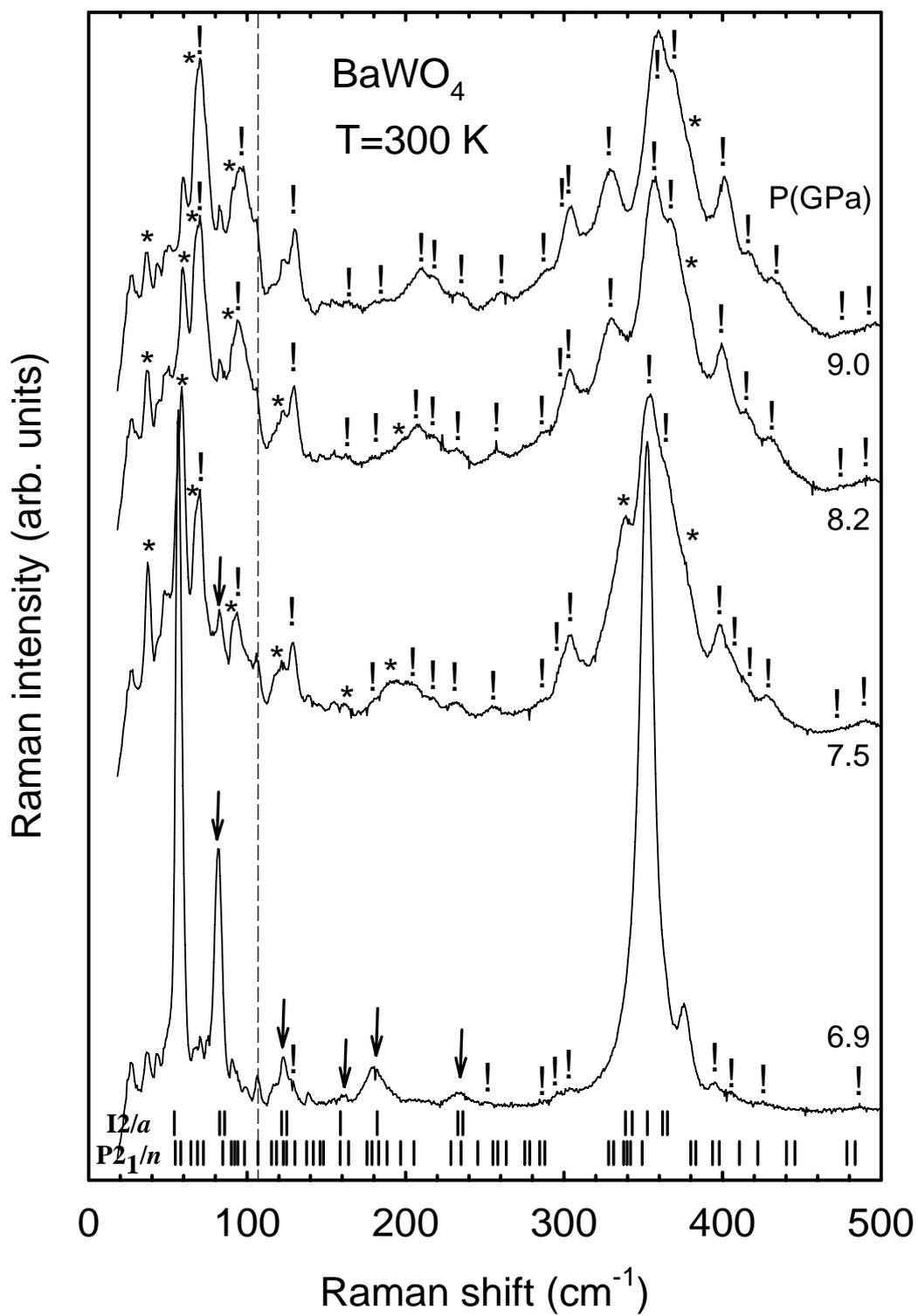

**Figure 5 (a).** F.J. Manjón et al.

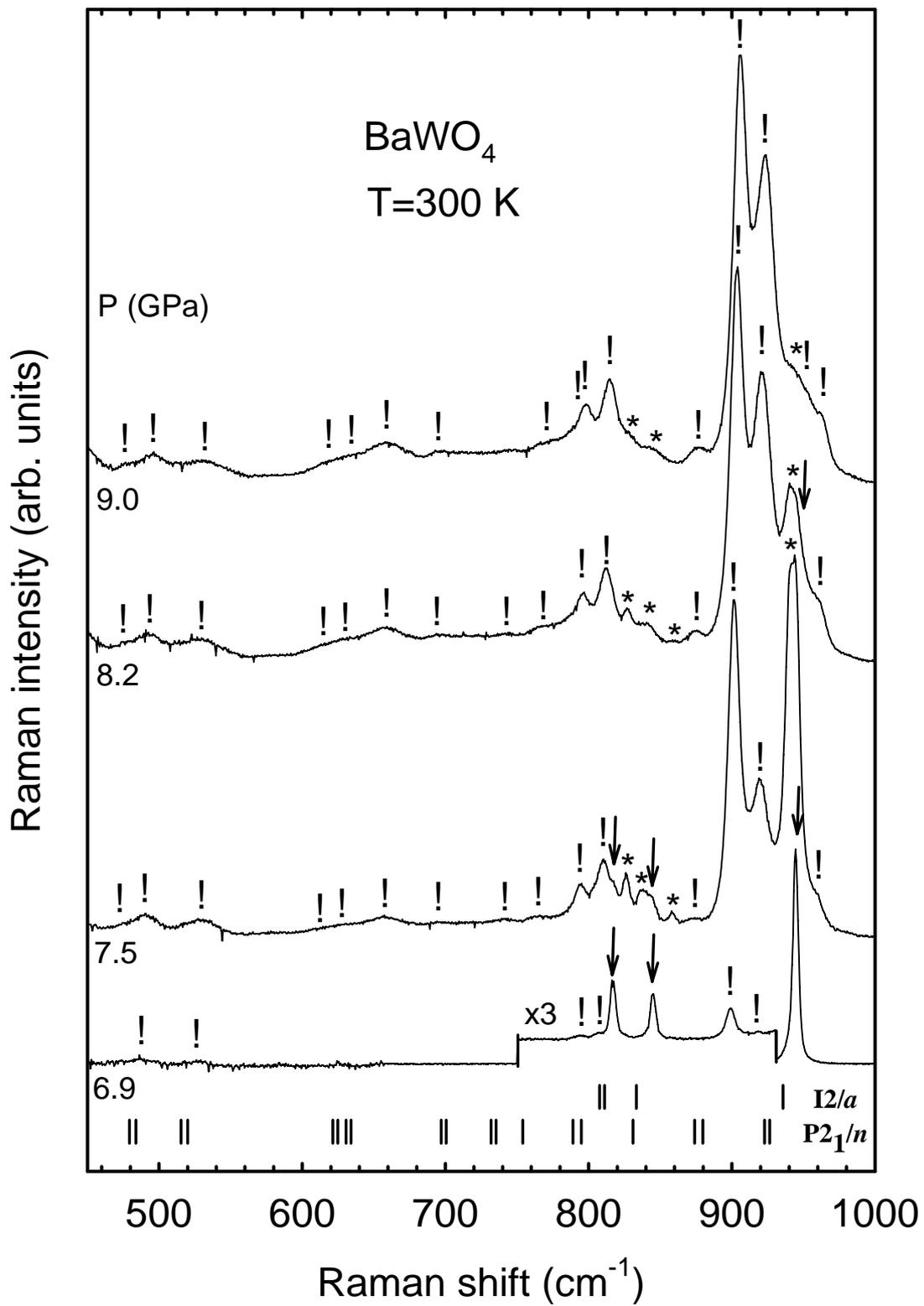

**Figure 5 (b).** F.J. Manjón et al.